\let\ifarxiv=\iftrue     
\pdfoutput=1

\ifarxiv

\documentclass[12pt,a4paper]{article}
\usepackage[a4paper,text={450pt,650pt},centering]{geometry}

\fi

\ifarxiv\else

\documentclass[11pt,a4paper]{article}
\usepackage{mathptmx}
\usepackage[a4paper,text={130mm,198mm}]{geometry}

\fi

\usepackage{amsmath,amssymb}
\usepackage[bookmarks=true,hyperfigures=true]{hyperref}
\usepackage{graphicx}
\usepackage[nosort]{cite}
\usepackage[bulletsep]{collref}

\usepackage{amsfonts}
\usepackage{latexsym,amsmath,amssymb}
\usepackage{color}
\usepackage{dsfont}

\usepackage{rotating}
\newlength{\arlength}

\usepackage{feynmp}

\usepackage{Dalgebra}

\newcommand{\col}{~,}
\newcommand{\pnt}{~.}
\newcommand{\AdS}{\text{AdS}}
\newcommand{\CFT}{\text{CFT}}

\newcommand{\YM}{\text{YM}}

\newcommand{\unitmatrix}{\mathds{1}}
\newcommand{\comm}[2]{\left[#1\smash[b]{\mathbin{,}}#2\right]}

\newcommand{\de}{\operatorname{d}\!}

\newcommand{\e}{\operatorname{e}}

\newcommand{\pthree}[3]{{}\{#1,#2,#3\}{}}

\newcommand{\pone}[1]{{}\{#1\}{}}

\newlength{\neglength}
\newlength{\diameter}

\newcommand{\svertex}[3][0.5]{%
\fmfiequ{#2}{point #1*length(#3) of #3}
}
\newcommand{\dvertex}[3]{%
\fmfiequ{#1}{point length(#3)/3 of #3}
\fmfiequ{#2}{point 2length(#3)/3 of #3}
}
\newcommand{\vvertex}[3]{%
\fmfipath{px}
\fmfiequ{px}{(0,ypart(#2))..(100,ypart(#2))}
\fmfiequ{#1}{point xpart(#3 intersectiontimes px) of #3}
}

\newcommand{\wigglywrap}[4]{%
\fmfipath{pi[]}
\fmfiset{pi1}{#1 ..controls (-0.175w,ypart(#1)) and (-0.175w,-0.15w) .. (xpart(vloc(__#2)),-0.15w)}
\fmfiset{pi2}{(xpart(vloc(__#2)),-0.15w) ..(xpart(vloc(__#3)),-0.15w)}

\fmfiset{pi3}{(xpart(vloc(__#3)),-0.15w) ..controls (1.175w,-0.15w) and (1.175w,ypart(#4)) .. #4}
\fmfi{photon}{pi1}
\fmfi{photon}{pi2}
\fmfi{photon}{pi3}
}

\newcommand{\chione}[1][black]{%
\fmftop{v1}
\fmfbottom{v3}
\fmfforce{(0.125w,h)}{v1}
\fmfforce{(0.125w,0)}{v3}
\fmffixed{(0.25w,0)}{v1,v2}
\fmffixed{(0.25w,0)}{v3,v4}
\fmf{plain,tension=0.5,right=0.25,fore=#1}{v1,vc1}
\fmf{plain,tension=0.5,left=0.25,fore=#1}{v2,vc1}
\fmf{plain,tension=1.25,fore=#1}{vc1,vc2}
\fmf{plain,tension=0.5,left=0.25,fore=#1}{v3,vc2}
\fmf{plain,tension=0.5,right=0.25,fore=#1}{v4,vc2}
\fmf{plain,tension=0.5,right=0,width=1mm,fore=#1}{v3,v4}
\fmfposition
\fmfipath{p[]}
\fmfipair{vd[],vm[],vu[]}
\fmfiset{p1}{vpath(__v1,__vc1)}
\fmfiset{p2}{vpath(__v2,__vc1)}
\fmfiset{p3}{vpath(__vc1,__vc2)}
\fmfiset{p4}{vpath(__v3,__vc2)}
\fmfiset{p5}{vpath(__v4,__vc2)}
\svertex{vm1}{p1}
\dvertex{vu1}{vd1}{p1}
\svertex{vm2}{p2}
\dvertex{vu2}{vd2}{p2}
\svertex{vm3}{p3}
\dvertex{vu3}{vd3}{p3}
\svertex{vm4}{p4}
\dvertex{vd4}{vu4}{p4}
\svertex{vm5}{p5}
\dvertex{vd5}{vu5}{p5}
}

\newcommand{\chioneg}[1][black]{%
\fmftop{v1}
\fmfbottom{v4}
\fmfforce{(0.125w,h)}{v1}
\fmfforce{(0.125w,0)}{v4}
\fmffixed{(0.25w,0)}{v1,v2}
\fmffixed{(0.25w,0)}{v2,v3}
\fmffixed{(0.25w,0)}{v4,v5}
\fmffixed{(0.25w,0)}{v5,v6}
\fmf{plain,tension=0.5,right=0.25,fore=#1}{v1,vc1}
\fmf{plain,tension=0.5,left=0.25,fore=#1}{v2,vc1}
  \fmf{plain,tension=1.25,fore=#1}{vc1,vc2}
\fmf{plain,tension=0.5,left=0.25,fore=#1}{v4,vc2}
\fmf{plain,tension=0.5,right=0.25,fore=#1}{v5,vc2}
\fmf{plain,fore=#1}{v3,v6}
\fmf{plain,tension=0.5,right=0,width=1mm,fore=#1}{v4,v6}
\fmfposition
\fmfipath{p[],pg}
\fmfipair{vd[],vm[],vu[],vgd[],vgm[],vgu[],vg[]}
\fmfiset{p1}{vpath(__v1,__vc1)}
\fmfiset{p2}{vpath(__v2,__vc1)}
\fmfiset{p3}{vpath(__vc1,__vc2)}
\fmfiset{p5}{vpath(__v5,__vc2)}
\fmfiset{p4}{vpath(__v4,__vc2)}
\fmfiset{pg}{vpath(__v3,__v6)}
\svertex{vm1}{p1}
\dvertex{vu1}{vd1}{p1}
\svertex{vm2}{p2}
\dvertex{vu2}{vd2}{p2}
\svertex{vm3}{p3}
\dvertex{vu3}{vd3}{p3}
\svertex{vm4}{p4}
\dvertex{vd4}{vu4}{p4}
\svertex{vm5}{p5}
\dvertex{vd5}{vu5}{p5}
\vvertex{vgu2}{vu2}{pg}
\vvertex{vgm2}{vm2}{pg}
\vvertex{vgd2}{vd2}{pg}
\vvertex{vgu3}{vu3}{pg}
\vvertex{vgm3}{vm3}{pg}
\vvertex{vgd3}{vd3}{pg}
\vvertex{vgu5}{vu5}{pg}
\vvertex{vgm5}{vm5}{pg}
\vvertex{vgd5}{vd5}{pg}
\vvertex{vg1}{vloc(__vc1)}{pg}
\vvertex{vg2}{vloc(__vc2)}{pg}
}

\newcommand{\chionetwo}[1][black]{%
\fmftop{v1}
\fmfbottom{v4}
\fmfforce{(0.125w,h)}{v1}
\fmfforce{(0.125w,0)}{v4}
\fmffixed{(0.25w,0)}{v1,v2}
\fmffixed{(0.25w,0)}{v2,v3}
\fmffixed{(0.25w,0)}{v4,v5}
\fmffixed{(0.25w,0)}{v5,v6}
\fmffixed{(0,whatever)}{vc1,vc3}
\fmffixed{(0,whatever)}{vc2,vc4}
\fmf{plain,tension=0.5,right=0.25}{v1,vc1}
\fmf{plain,tension=0.5,left=0.25}{v2,vc1}
\fmf{phantom,tension=0.5,right=0.25}{v2,vc2}
\fmf{plain,tension=0.5,left=0.25}{v3,vc2}
\fmf{plain,tension=0.5,left=0.25}{v4,vc3}
\fmf{phantom,tension=0.5,right=0.25}{v5,vc3}
\fmf{plain,tension=0.5,left=0.25}{v5,vc4}
\fmf{plain,tension=0.5,right=0.25}{v6,vc4}
\fmf{plain,tension=1.25,left=0}{vc1,vc3}
\fmf{plain,tension=1.25,left=0}{vc2,vc4}
\fmffreeze
\fmf{plain,tension=1,left=0}{vc2,vc3}
\fmf{plain,tension=0.5,right=0,width=1mm}{v4,v6}
\fmffreeze
\fmfposition
\fmfipath{p[]}
\fmfipair{vd[],vm[],vu[]}
\fmfiset{p1}{vpath(__v1,__vc1)}
\fmfiset{p2}{vpath(__v2,__vc1)}
\fmfiset{p6}{vpath(__v3,__vc2)}
\fmfiset{p4}{vpath(__v4,__vc3)}
\fmfiset{p8}{vpath(__v5,__vc4)}
\fmfiset{p9}{vpath(__v6,__vc4)}
\fmfiset{p3}{vpath(__vc1,__vc3)}
\fmfiset{p7}{vpath(__vc2,__vc4)}
\fmfiset{p5}{vpath(__vc2,__vc3)}
\svertex{vm1}{p1}
\svertex{vm2}{p2}
\svertex{vm3}{p3}
\svertex{vm4}{p4}
\svertex{vm5}{p5}
\svertex{vm6}{p6}
\svertex{vm7}{p7}
\svertex{vm8}{p8}
\svertex{vm9}{p9}
}

\newcommand{\chionetwoone}[1][black]{%
\fmftop{v1}
\fmfbottom{v4}
\fmfforce{(0.125w,h)}{v1}
\fmfforce{(0.125w,0)}{v4}
\fmffixed{(0.25w,0)}{v1,v2}
\fmffixed{(0.25w,0)}{v2,v3}
\fmffixed{(0.25w,0)}{v4,v5}
\fmffixed{(0.25w,0)}{v5,v6}
\fmffixed{(0,whatever)}{vc1,vc3}
\fmffixed{(0,whatever)}{vb2,vb4}
\fmffixed{(0,whatever)}{vc1,vb1}
\fmffixed{(0,whatever)}{vc1,vb3}
\fmffixed{(whatever,0)}{vb1,vb2}
\fmffixed{(whatever,0)}{vb3,vb4}
\fmf{plain,tension=0.5,right=0.25}{v1,vc1}
\fmf{plain,tension=0.5,left=0.25}{v2,vc1}
\fmf{phantom,tension=0.5,right=0.25}{v2,vb2}
\fmf{plain,tension=0.5,left=0.25}{v3,vb2}
\fmf{plain,tension=0.5,left=0.25}{v4,vc3}
\fmf{plain,tension=0.5,right=0.25}{v5,vc3}
\fmf{phantom,tension=0.5,left=0.25}{v5,vb4}
\fmf{plain,tension=0.5,right=0.25}{v6,vb4}
\fmf{plain,tension=1.25,left=0}{vc1,vb1}
\fmf{plain,tension=1.25,left=0}{vb1,vb3}
\fmf{plain,tension=1.25,left=0}{vb3,vc3}
\fmf{plain,tension=1.25,left=0}{vb2,vb4}
\fmffreeze
\fmf{plain,tension=1,left=0}{vb1,vb2}
\fmf{plain,tension=1,left=0}{vb3,vb4}
\fmf{plain,tension=0.5,right=0,width=1mm}{v4,v6}
\fmffreeze
\fmfposition
}

\newcommand{\chionetwothree}[1][black]{%
\fmftop{v1}
\fmfbottom{v5}
\fmfforce{(0.125w,h)}{v1}
\fmfforce{(0.125w,0)}{v5}
\fmffixed{(0.25w,0)}{v1,v2}
\fmffixed{(0.25w,0)}{v2,v3}
\fmffixed{(0.25w,0)}{v3,v4}
\fmffixed{(0.25w,0)}{v5,v6}
\fmffixed{(0.25w,0)}{v6,v7}
\fmffixed{(0.25w,0)}{v7,v8}
\fmffixed{(0,whatever)}{vc1,vc4}
\fmffixed{(0,whatever)}{vc2,vc5}
\fmffixed{(0,whatever)}{vc3,vc6}

\fmf{plain,tension=0.5,right=0.25}{v1,vc1}
\fmf{plain,tension=0.5,left=0.25}{v2,vc1}
\fmf{phantom,tension=0.5,right=0.25}{v2,vc2}
\fmf{plain,tension=0.5,left=0.25}{v3,vc2}
\fmf{phantom,tension=0.5,right=0.25}{v3,vc3}
\fmf{plain,tension=0.5,left=0.25}{v4,vc3}
\fmf{plain,tension=0.5,left=0.25}{v5,vc4}
\fmf{phantom,tension=0.5,right=0.25}{v6,vc4}
\fmf{plain,tension=0.5,left=0.25}{v6,vc5}
\fmf{phantom,tension=0.5,right=0.25}{v7,vc5}
\fmf{plain,tension=0.5,left=0.25}{v7,vc6}
\fmf{plain,tension=0.5,right=0.25}{v8,vc6}
\fmf{plain,tension=1.25,left=0}{vc1,vc4}
\fmf{plain,tension=1.25,left=0}{vc2,vc5}
\fmf{plain,tension=1.25,left=0}{vc3,vc6}
\fmffreeze
\fmf{plain,tension=1,left=0}{vc4,vc2}
\fmf{plain,tension=1,left=0}{vc5,vc3}
\fmf{plain,tension=0.5,right=0,width=1mm}{v5,v8}
\fmffreeze
\fmfposition
}

\newcommand{\chitwoonethree}[1][black]{%
\fmftop{v1}
\fmfbottom{v5}
\fmfforce{(0.125w,h)}{v1}
\fmfforce{(0.125w,0)}{v5}
\fmffixed{(0.25w,0)}{v1,v2}
\fmffixed{(0.25w,0)}{v2,v3}
\fmffixed{(0.25w,0)}{v3,v4}
\fmffixed{(0.25w,0)}{v5,v6}
\fmffixed{(0.25w,0)}{v6,v7}
\fmffixed{(0.25w,0)}{v7,v8}
\fmffixed{(whatever,0.5h)}{v5,vc1}
\fmffixed{(0,whatever)}{vc1,vc4}
\fmffixed{(0,whatever)}{vc2,vc5}
\fmffixed{(0,whatever)}{vc3,vc6}
\fmffixed{(whatever,0)}{vc1,vc3}
\fmffixed{(whatever,0)}{vc3,vc5}

\fmf{plain,tension=0.5,right=0.125}{v1,vc1}
\fmf{phantom,tension=0.5,left=0.25}{v2,vc1}
\fmf{plain,tension=0.5,right=0.25}{v2,vc2}
\fmf{plain,tension=0.5,left=0.25}{v3,vc2}
\fmf{phantom,tension=0.5,right=0.25}{v3,vc3}
\fmf{plain,tension=0.5,left=0.125}{v4,vc3}
\fmf{plain,tension=0.5,left=0.25}{v5,vc4}
\fmf{plain,tension=0.5,right=0.25}{v6,vc4}
\fmf{phantom,tension=0.5,left=0.25}{v6,vc5}
\fmf{phantom,tension=0.5,right=0.25}{v7,vc5}
\fmf{plain,tension=0.5,left=0.25}{v7,vc6}
\fmf{plain,tension=0.5,right=0.25}{v8,vc6}
\fmf{plain,tension=1.25,left=0}{vc1,vc4}
\fmf{plain,tension=1.25,left=0}{vc2,vc5}
\fmf{plain,tension=1.25,left=0}{vc3,vc6}
\fmffreeze
\fmf{plain,tension=1,left=0}{vc1,vc5}
\fmf{plain,tension=1,left=0}{vc5,vc3}
\fmf{plain,tension=0.5,right=0,width=1mm}{v5,v8}
\fmffreeze
\fmfposition
}

\newcommand{\chionethreetwo}[1][black]{%
\fmftop{v1}
\fmfbottom{v5}
\fmfforce{(0.125w,h)}{v1}
\fmfforce{(0.125w,0)}{v5}
\fmffixed{(0.25w,0)}{v1,v2}
\fmffixed{(0.25w,0)}{v2,v3}
\fmffixed{(0.25w,0)}{v3,v4}
\fmffixed{(0.25w,0)}{v5,v6}
\fmffixed{(0.25w,0)}{v6,v7}
\fmffixed{(0.25w,0)}{v7,v8}
\fmffixed{(whatever,0.5h)}{v5,vc2}
\fmffixed{(0,whatever)}{vc1,vc4}
\fmffixed{(0,whatever)}{vc2,vc5}
\fmffixed{(0,whatever)}{vc3,vc6}
\fmffixed{(whatever,0)}{vc2,vc4}
\fmffixed{(whatever,0)}{vc4,vc6}
\fmf{plain,tension=0.5,right=0.25}{v1,vc1}
\fmf{plain,tension=0.5,left=0.25}{v2,vc1}
\fmf{phantom,tension=0.5,right=0.25}{v2,vc2}
\fmf{phantom,tension=0.5,left=0.25}{v3,vc2}
\fmf{plain,tension=0.5,right=0.25}{v3,vc3}
\fmf{plain,tension=0.5,left=0.25}{v4,vc3}
\fmf{plain,tension=0.5,left=0.125}{v5,vc4}
\fmf{phantom,tension=0.5,right=0.25}{v6,vc4}
\fmf{plain,tension=0.5,left=0.25}{v6,vc5}
\fmf{plain,tension=0.5,right=0.25}{v7,vc5}
\fmf{phantom,tension=0.5,left=0.25}{v7,vc6}
\fmf{plain,tension=0.5,right=0.125}{v8,vc6}
\fmf{plain,tension=1.25,left=0}{vc1,vc4}
\fmf{plain,tension=1.25,left=0}{vc2,vc5}
\fmf{plain,tension=1.25,left=0}{vc3,vc6}
\fmffreeze
\fmf{plain,tension=1,left=0}{vc2,vc4}
\fmf{plain,tension=1,left=0}{vc2,vc6}
\fmf{plain,tension=0.5,right=0,width=1mm}{v5,v8}
\fmffreeze
\fmfposition
}

\newcommand{\idzero}[1][black]{%
\fmftop{v1}
\fmfbottom{v2}
\fmfforce{(0.5w,h)}{v1}
\fmfforce{(0.5w,0)}{v2}
\fmffixed{(0.01w,0)}{v2l,v2}
\fmffixed{(0.01w,0)}{v2,v2r}
\fmf{plain,fore=#1}{v1,v2}
\fmf{plain,tension=0.5,right=0,width=1mm,fore=#1}{v2l,v2r}
\fmfposition
\fmfipath{p[]}
\fmfiset{p1}{vpath(__v1,__v2)}
}

\newcommand{\idone}[1][black]{%
\fmftop{v1}
\fmfbottom{v3}
\fmfforce{(0.125w,h)}{v1}
\fmfforce{(0.125w,0)}{v3}
\fmffixed{(0.25w,0)}{v1,v2}
\fmffixed{(0.25w,0)}{v3,v4}
\fmf{plain,fore=#1}{v1,v3}
\fmf{phantom}{v1,vc1}
\fmf{phantom}{vc1,v3}
\fmf{plain,fore=#1}{v2,v4}
\fmf{plain,tension=0.5,right=0,width=1mm,fore=#1}{v3,v4}
\fmfposition
\fmfipath{p[]}
\fmfiset{p1}{vpath(__v1,__v3)}

\fmfiset{p2}{vpath(__v2,__v4)}
}

\newcommand{\idtwo}[1][black]{%
\fmftop{v1}
\fmfbottom{v4}
\fmfforce{(0.125w,h)}{v1}
\fmfforce{(0.125w,0)}{v4}
\fmffixed{(0.25w,0)}{v1,v2}
\fmffixed{(0.25w,0)}{v2,v3}
\fmffixed{(0.25w,0)}{v4,v5}
\fmffixed{(0.25w,0)}{v5,v6}
\fmf{plain,fore=#1}{v1,v4}
\fmf{plain,fore=#1}{v2,v5}
\fmf{plain,fore=#1}{v3,v6}
\fmf{plain,tension=0.5,right=0,width=1mm,fore=#1}{v4,v6}
\fmfposition
\fmfipath{p[]}
\fmfiset{p1}{vpath(__v1,__v4)}
\fmfiset{p2}{vpath(__v2,__v5)}
\fmfiset{p3}{vpath(__v3,__v6)}
}

\newcommand{\vacpol}[3][black]{%
\fmfcmd{
begingroup;
save t,v,tv,do,di,ppol,pstr,dia;
path ppol,pstr;
pair v[],tv[],do[],di[];
ppol=vpath(__#2,__#3);
t1=arctime (1/3*arclength ppol) of ppol;
t2=arctime (2/3*arclength ppol) of ppol;
v1=point t1 of ppol;
v2=point t2 of ppol;
pstr=v1--v2;
t3=arctime (0.5*arclength pstr) of pstr;
v3=point t3 of pstr;
dia=arclength pstr; 
fill(fullcircle scaled dia shifted v3) withcolor #1;
endgroup;
}
}

\newcommand{\cVat}[7][]{%
\settoheight{\eqoff}{$\times$}%
\setlength{\eqoff}{0.5\eqoff}%
\addtolength{\eqoff}{-13\unitlength}%
\raisebox{\eqoff}{%
\fmfframe(2,1)(2,1){%
\begin{fmfchar*}(21,24)
\fmftop{v3}
\fmfbottom{v2}
\fmfforce{(w,h)}{v3}
\fmfforce{(w,0)}{v2}
\fmfpoly{phantom}{v1,v2,v3}
\fmf{phantom,tension=2}{v1,vg1}
\fmf{phantom}{v2,vg1}
\fmf{phantom}{v3,vg1}
\fmffreeze
\fmf{plain,tension=1}{vg2,v2}
\fmf{plain,tension=1}{vg3,v3}
\fmf{phantom}{vg1,v1}
\fmf{phantom,tension=0.5}{vg1,vg2}
\fmf{phantom,tension=0.5}{vg1,vg3}
\fmffreeze
\fmfposition
\fmf{#5}{vg2,vg3}
\fmfipath{p[],pca}
\fmfipair{vm[],vo[],vi[]}
\fmfiset{p1}{vpath(__vg1,__v1)}
\fmfiset{p2}{vpath(__vg2,__v2)}
\fmfiset{p3}{vpath(__vg3,__v3)}
\fmfiset{p4}{vpath(__vg2,__vg3)}
\fmfiset{p5}{vpath(__vg3,__vg1)}
\fmfiset{p6}{vpath(__vg1,__vg2)}
{#1}
\fmfis{#2,ptext.clen=7,ptext.hin=3,ptext.hout=3,ptext.oin=6,ptext.oout=6,ptext.sep=;}{reverse(p1)}
\fmfis{#3,ptext.clen=7,ptext.hin=-10,ptext.hout=-10,ptext.oin=6,ptext.oout=6,ptext.sep=;}{p6}
\fmfis{#4,ptext.clen=7,ptext.hin=3,ptext.hout=3,ptext.oin=6,ptext.oout=6,ptext.sep=;}{p5}
\fmfis{#6,ptext.clen=7,ptext.hin=3,ptext.hout=3,ptext.oin=6,ptext.oout=6,ptext.sep=;}{p2}
\fmfis{#7,ptext.clen=7,ptext.hin=-10,ptext.hout=-10,ptext.oin=6,ptext.oout=6,ptext.sep=;}{p3}
\end{fmfchar*}}}
}

\DeclareMathOperator{\tr}{tr}

\DeclareMathOperator{\res}{Res}

\DeclareMathOperator{\perm}{P}
\DeclareMathOperator{\Kop}{K}
\DeclareMathOperator{\Rop}{R}
\DeclareMathOperator{\D}{D}
\DeclareMathOperator{\barD}{\vphantom{\D}\smash[t]{\bar{\mathrm{D}}}}

\newlength{\eqoff}
\newlength{\eqofftwo}

\newlength{\unit}
\newlength{\linew}
\let\oldbfseries=\bfseries
\let\oldmdseries=\mdseries
\let\oldnormalfont=\normalfont
\renewcommand{\bfseries}{\oldbfseries\boldmath}
\renewcommand{\mdseries}{\oldmdseries\unboldmath}
\renewcommand{\normalfont}{\oldnormalfont\unboldmath}

\allowdisplaybreaks[3]

\numberwithin{equation}{section}

\usepackage[font=small,labelfont=bf,width=0.85\textwidth]{caption}

\providecommand{\hypersetup}[1]{}
\providecommand{\texorpdfstring}[2]{#1}

\hypersetup{plainpages=false}
\hypersetup{pdfpagemode=UseNone}
\hypersetup{bookmarksnumbered=true}
\hypersetup{pdfstartview=FitH}
\hypersetup{colorlinks=false}
\hypersetup{citebordercolor={.5 1 .5}}
\hypersetup{urlbordercolor={.5 1 1}}
\hypersetup{linkbordercolor={1 .7 .7}}

\providecommand{\arxivref}[2]{\href{http://arxiv.org/abs/#1}{#2}}
\providecommand{\doiref}[2]{\href{http://dx.doi.org/#1}{#2}}

\providecommand{\href}[2]{#2}
\providecommand{\arxivlink}[1]{\href{http://arxiv.org/abs/#1}{arxiv:#1}}

\newcommand{\mympostgrey}{0.75 white}

\newcommand{\idRone}[4][]{\settoheight{\eqoff}{$\times$}%
\setlength{\eqoff}{0.5\eqoff}%
\addtolength{\eqoff}{-12\unitlength}%
\raisebox{\eqoff}{%
\fmfframe(-10,2)(-10,2){%
\begin{fmfchar*}(20,20)
\idzero[(#2,,#3,,#4)]
\vacpol[(#2,#3,#4)]{v1}{v2}
\fmfipair{v[]}
\svertex{v3}{vpath(__v1,__v2)}
\fmfiv{plain,label=$\textcolor{white}{#1}$,l.dist=0}{v3}
\end{fmfchar*}}}
}
\newcommand{\idRtwo}[3]{\settoheight{\eqoff}{$\times$}%
\setlength{\eqoff}{0.5\eqoff}%
\addtolength{\eqoff}{-12\unitlength}%
\raisebox{\eqoff}{%
\fmfframe(-0.5,2)(-10.5,2){%
\begin{fmfchar*}(20,20)
\idone[(#1,,#2,,#3)]
\fmfipair{g[]}
\svertex{g1}{p1}
\svertex{g2}{p2}
\fmfi{wiggly,fore=(#1,,#2,,#3)}{g1{dir -60}..{dir 60}g2}
\fmfi{wiggly,fore=(#1,,#2,,#3)}{g1{dir 60}..{dir -60}g2}
\end{fmfchar*}}}
\settoheight{\eqoff}{$\times$}%
\setlength{\eqoff}{0.5\eqoff}%
\addtolength{\eqoff}{-12\unitlength}%
\raisebox{\eqoff}{%
\fmfframe(-0.50,2)(-10.5,2){%
\begin{fmfchar*}(20,20)
\idone[(#1,,#2,,#3)]
\fmfipair{g[],gu[],gd[]}
\svertex{g1}{p1}
\dvertex{gu2}{gd2}{p2}
\fmfi{wiggly,fore=(#1,,#2,,#3)}{g1--gu2}
\fmfi{wiggly,fore=(#1,,#2,,#3)}{g1--gd2}
\end{fmfchar*}}}
\settoheight{\eqoff}{$\times$}%
\setlength{\eqoff}{0.5\eqoff}%
\addtolength{\eqoff}{-12\unitlength}%
\raisebox{\eqoff}{%
\fmfframe(-0.5,2)(-10.5,2){%
\begin{fmfchar*}(20,20)
\idone[(#1,,#2,,#3)]
\fmfipair{g[],gu[],gd[]}
\svertex{g1}{p1}
\svertex{g2}{p2}
\dvertex{gu2}{gd2}{p2}
\fmfi{wiggly,fore=(#1,,#2,,#3)}{g1{dir 60}..gu2}
\fmfi{wiggly,fore=(#1,,#2,,#3)}{gu2{dir 210}..{dir -30}gd2}
\end{fmfchar*}}}
\settoheight{\eqoff}{$\times$}%
\setlength{\eqoff}{0.5\eqoff}%
\addtolength{\eqoff}{-12\unitlength}%
\raisebox{\eqoff}{%
\fmfframe(-0.5,2)(-10.5,2){%
\begin{fmfchar*}(20,20)
\idone[(#1,,#2,,#3)]
\fmfipair{g[],gu[],gd[]}
\svertex{g1}{p1}
\svertex{g2}{p2}
\dvertex{gu2}{gd2}{p2}
\fmfi{wiggly,fore=(#1,,#2,,#3)}{g1{dir 60}..gu2}
\fmfi{wiggly,fore=(#1,,#2,,#3)}{gu2{dir -30}..{dir 210}gd2}
\end{fmfchar*}}}
\settoheight{\eqoff}{$\times$}%
\setlength{\eqoff}{0.5\eqoff}%
\addtolength{\eqoff}{-11\unitlength}%
\raisebox{\eqoff}{%
\fmfframe(-0.5,1)(-10.5,1){%
\begin{fmfchar*}(20,20)
\idone[(#1,,#2,,#3)]
\fmfipair{g[],gu[],gd[]}
\svertex{g1}{p1}
\svertex{g2}{p2}
\dvertex{gu2}{gd2}{p2}
\fmfi{wiggly,fore=(#1,,#2,,#3)}{g1{dir 0}..{dir 0}g2}
\fmfi{wiggly,fore=(#1,,#2,,#3)}{gu2{dir -30}..{dir 210}gd2}
\end{fmfchar*}}}
\settoheight{\eqoff}{$\times$}%
\setlength{\eqoff}{0.5\eqoff}%
\addtolength{\eqoff}{-12\unitlength}%
\raisebox{\eqoff}{%
\fmfframe(-0.5,2)(-10.5,2){%
\begin{fmfchar*}(20,20)
\idone[(#1,,#2,,#3)]
\fmfipair{gu[],gd[]}
\dvertex{gu1}{gd1}{p1}
\dvertex{gu2}{gd2}{p2}
\fmfi{wiggly,fore=(#1,,#2,,#3)}{gu1{dir 0}..{dir 0}gu2}
\fmfi{wiggly,fore=(#1,,#2,,#3)}{gd1{dir 0}..{dir 0}gd2}
\end{fmfchar*}}}
\settoheight{\eqoff}{$\times$}%
\setlength{\eqoff}{0.5\eqoff}%
\addtolength{\eqoff}{-12\unitlength}%
\raisebox{\eqoff}{%
\fmfframe(-0.5,2)(-10.5,2){%
\begin{fmfchar*}(20,20)
\idone[(#1,,#2,,#3)]
\fmf{phantom}{v1,v0}
\fmf{phantom}{v2,v0}
\fmf{phantom}{v3,v0}
\fmf{phantom}{v4,v0}
\fmffreeze
\fmfposition
\fmfipair{gu[],gd[],g[]}
\dvertex{gu1}{gd1}{p1}
\svertex{g1}{p2}
\fmfi{wiggly,fore=(#1,,#2,,#3)}{gu1--vloc(__v0)}
\fmfi{wiggly,fore=(#1,,#2,,#3)}{gd1--vloc(__v0)}
\fmfi{wiggly,fore=(#1,,#2,,#3)}{g1--vloc(__v0)}
\end{fmfchar*}}}
}

\newcommand{\idRthree}[3]{
\settoheight{\eqoff}{$\times$}%
\setlength{\eqoff}{0.5\eqoff}%
\addtolength{\eqoff}{-11\unitlength}%
\raisebox{\eqoff}{%
\fmfframe(-0.5,1)(-5.5,1){%
\begin{fmfchar*}(20,20)
\idtwo[(#1,,#2,,#3)]
\fmfipair{g[]}
\svertex{g1}{p1}
\svertex{g2}{p2}
\svertex{g3}{p3}
\fmfi{wiggly,fore=(#1,,#2,,#3)}{g1{dir 0}..{dir 0}g2}
\fmfi{wiggly,fore=(#1,,#2,,#3)}{g2{dir 0}..{dir 0}g3}
\end{fmfchar*}}}
\settoheight{\eqoff}{$\times$}%
\setlength{\eqoff}{0.5\eqoff}%
\addtolength{\eqoff}{-11\unitlength}%
\raisebox{\eqoff}{%
\fmfframe(-0.5,1)(-5.5,1){%
\begin{fmfchar*}(20,20)
\idtwo[(#1,,#2,,#3)]
\fmfipair{gu[],gd[]}
\dvertex{gu1}{gd1}{p1}
\dvertex{gu2}{gd2}{p2}
\dvertex{gu3}{gd3}{p3}
\fmfi{wiggly,fore=(#1,,#2,,#3)}{gu1{dir 0}..{dir 0}gu2}
\fmfi{wiggly,fore=(#1,,#2,,#3)}{gd2{dir 0}..{dir 0}gd3}
\end{fmfchar*}}}
}

\newcommand{\chioneRtwo}[3]{
\settoheight{\eqoff}{$\times$}%
\setlength{\eqoff}{0.5\eqoff}%
\addtolength{\eqoff}{-12\unitlength}%
\raisebox{\eqoff}{%
\fmfframe(-0.5,2)(-10.5,2){%
\begin{fmfchar*}(20,20)
\chione[(#1,,#2,,#3)]
\fmfi{wiggly,fore=(#1,,#2,,#3)}{vm2{dir -45}..{dir -135}vm5}
\end{fmfchar*}}}
\settoheight{\eqoff}{$\times$}%
\setlength{\eqoff}{0.5\eqoff}%
\addtolength{\eqoff}{-12\unitlength}%
\raisebox{\eqoff}{%
\fmfframe(-0.5,2)(-10.5,2){%
\begin{fmfchar*}(20,20)
\chione
\fmfi{wiggly}{vm3{dir 0}..{dir -135}vm5}
\end{fmfchar*}}}
\settoheight{\eqoff}{$\times$}%
\setlength{\eqoff}{0.5\eqoff}%
\addtolength{\eqoff}{-12\unitlength}%
\raisebox{\eqoff}{%
\fmfframe(-0.5,2)(-10.5,2){%
\begin{fmfchar*}(20,20)
\chione
\fmfi{wiggly}{vm4--vm5}
\end{fmfchar*}}}
}

\newcommand{\chioneRthree}[3]{
\settoheight{\eqoff}{$\times$}%
\setlength{\eqoff}{0.5\eqoff}%
\addtolength{\eqoff}{-12\unitlength}%
\raisebox{\eqoff}{%
\fmfframe(-0.5,2)(-5.5,2){%
\begin{fmfchar*}(20,20)
\chioneg[(#1,,#2,,#3)]
\fmfi{wiggly,fore=(#1,,#2,,#3)}{vm2--vgm2}
\end{fmfchar*}}}
\settoheight{\eqoff}{$\times$}%
\setlength{\eqoff}{0.5\eqoff}%
\addtolength{\eqoff}{-12\unitlength}%
\raisebox{\eqoff}{%
\fmfframe(-0.5,2)(-5.5,2){%
\begin{fmfchar*}(20,20)
\chioneg[(#1,,#2,,#3)]
\fmfi{wiggly,fore=(#1,,#2,,#3)}{vm3--vgm3}
\end{fmfchar*}}}
\settoheight{\eqoff}{$\times$}%
\setlength{\eqoff}{0.5\eqoff}%
\addtolength{\eqoff}{-12\unitlength}%
\raisebox{\eqoff}{%
\fmfframe(-0.5,2)(-5.5,2){%
\begin{fmfchar*}(20,20)
\chioneg[(#1,,#2,,#3)]
\fmfi{wiggly,fore=(#1,,#2,,#3)}{vm5--vgm5}
\end{fmfchar*}}}
}

\newcommand{\chionetwoRthree}[3]{
\settoheight{\eqoff}{$\times$}%
\setlength{\eqoff}{0.5\eqoff}%
\addtolength{\eqoff}{-12\unitlength}%
\raisebox{\eqoff}{%
\fmfframe(-0.5,2)(-5.5,2){%
\begin{fmfchar*}(20,20)
\chionetwo[(#1,,#2,,#3)]
\end{fmfchar*}}}
}

\begin{document}
\begin{fmffile}{fullgraphs}

\fmfcmd{%
input Dalgebra
def getmid(suffix p) =
  pair p.mid[], p.off[], p.dir[];
  for i=0 upto 36:
    p.dir[i] = dir(5*i);
    p.mid[i]+p.off[i] = directionpoint p.dir[i] of p;
    p.mid[i]-p.off[i] = directionpoint -p.dir[i] of p;
  endfor
enddef;
}

\fmfcmd{%
marksize=2mm;
def draw_mark(expr p,a) =
  begingroup
    save t,tip,dma,dmb; pair tip,dma,dmb;
    t=arctime a of p;
    tip =marksize*unitvector direction t of p;
    dma =marksize*unitvector direction t of p rotated -45;
    dmb =marksize*unitvector direction t of p rotated 45;
    linejoin:=beveled;
    draw (-.5dma.. .5tip-- -.5dmb) shifted point t of p;
  endgroup
enddef;
style_def derplain expr p =
    save amid;
    amid=.5*arclength p;
    draw_mark(p, amid);
    draw p;
enddef;
def draw_marks(expr p,a) =
  begingroup
    save t,tip,dma,dmb,dmo; pair tip,dma,dmb,dmo;
    t=arctime a of p;
    tip =marksize*unitvector direction t of p;
    dma =marksize*unitvector direction t of p rotated -45;
    dmb =marksize*unitvector direction t of p rotated 45;
    dmo =marksize*unitvector direction t of p rotated 90;
    linejoin:=beveled;
    draw (-.5dma.. .5tip-- -.5dmb) shifted point t of p withcolor 0white;
    draw (-.5dmo.. .5dmo) shifted point t of p;
  endgroup
enddef;
style_def derplains expr p =
    save amid;
    amid=.5*arclength p;
    draw_marks(p, amid);
    draw p;
enddef;
def draw_markss(expr p,a) =
  begingroup
    save t,tip,dma,dmb,dmo; pair tip,dma,dmb,dmo;
    t=arctime a of p;
    tip =marksize*unitvector direction t of p;
    dma =marksize*unitvector direction t of p rotated -45;
    dmb =marksize*unitvector direction t of p rotated 45;
    dmo =marksize*unitvector direction t of p rotated 90;
    linejoin:=beveled;
    draw (-.5dma.. .5tip-- -.5dmb) shifted point t of p withcolor 0white;
    draw (-.5dmo.. .5dmo) shifted point arctime a+0.25 mm of p of p;
    draw (-.5dmo.. .5dmo) shifted point arctime a-0.25 mm of p of p;
  endgroup
enddef;
style_def derplainss expr p =
    save amid;
    amid=.5*arclength p;
    draw_markss(p, amid);
    draw p;
enddef;
style_def dblderplains expr p =
    save amidm;
    save amidp;
    amidm=.5*arclength p-0.75mm;
    amidp=.5*arclength p+0.75mm;
    draw_mark(p, amidm);
    draw_marks(p, amidp);
    draw p;
enddef;
style_def dblderplainss expr p =
    save amidm;
    save amidp;
    amidm=.5*arclength p-0.75mm;
    amidp=.5*arclength p+0.75mm;
    draw_mark(p, amidm);
    draw_markss(p, amidp);
    draw p;
enddef;
style_def dblderplainsss expr p =
    save amidm;
    save amidp;
    amidm=.5*arclength p-0.75mm;
    amidp=.5*arclength p+0.75mm;
    draw_marks(p, amidm);
    draw_markss(p, amidp);
    draw p;
enddef;
}

\fmfcmd{%
thin := 1pt; 
thick := 2thin;
arrow_len := 4mm;
arrow_ang := 15;
curly_len := 3mm;
dash_len := 1.5mm; 
dot_len := 1mm; 
wiggly_len := 2mm; 
wiggly_slope := 60;
zigzag_len := 2mm;
zigzag_width := 2thick;
decor_size := 5mm;
dot_size := 2thick;
}


\thispagestyle{empty}
\phantomsection
\addcontentsline{toc}{section}{Title}

\begin{flushright}\footnotesize%
\texttt{HU-MATH-2010-23},
\texttt{HU-EP-10/88},
\texttt{\arxivlink{1012.3984}}\\
overview article: \texttt{\arxivlink{1012.3982}}%
\vspace{1em}%
\end{flushright}

\begingroup\parindent0pt
\begingroup\bfseries\ifarxiv\Large\else\LARGE\fi
\hypersetup{pdftitle={Review of AdS/CFT Integrability, Chapter I.2: The spectrum from perturbative gauge theory}}%
Review of AdS/CFT Integrability, Chapter I.2: \\
The spectrum from perturbative gauge theory
\par\endgroup
\vspace{1.5em}
\begingroup\ifarxiv\scshape\else\large\fi%
\hypersetup{pdfauthor={C. Sieg}}%
C.\ Sieg
\par\endgroup
\vspace{1em}
\begingroup\itshape
Institut f\"ur Mathematik und Institut f\"ur Physik,
Humboldt-Universit\"at zu Berlin \\
Johann von Neumann Haus, Rudower Chaussee 25,
12489 Berlin,
Germany
\\[0.5\baselineskip]
Niels Bohr International Academy,
Niels Bohr Institute \\
Blegdamsvej 17,
2100 Copenhagen,
Denmark
\par\endgroup
\vspace{1em}
\begingroup\ttfamily
csieg@math.hu-berlin.de
\par\endgroup
\vspace{1.0em}
\endgroup

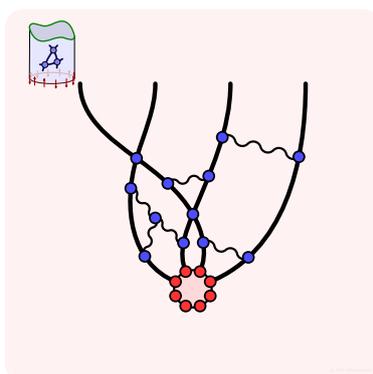
\begin{center}
\includegraphics[width=5cm]{TitleI2.mps}
\vspace{1.0em}
\end{center}

\paragraph{Abstract:}
We review the 
constructions and tests of the dilatation operator and
of the spectrum of composite operators 
in the flavour $SU(2)$ subsector of $\mathcal{N}=4$ SYM 
in the planar limit by explicit Feynman graph calculations with emphasis
on analyses beyond one loop.
From four loops on, the dilatation operator determines the spectrum only 
in the asymptotic regime, i.e.\ to a loop order which is strictly smaller 
than the number of elementary fields of the composite operators.
We review also the calculations which take a first step beyond this limitation
by including the leading wrapping corrections.

\ifarxiv\else
\fi
\hypersetup{pdfsubject={MSC (2010): 81T18; 81T13; 81T60}}%

\ifarxiv\else

\fi
\hypersetup{pdfkeywords={Feynman diagrams; Supersymmetric Yang-Mills theory; Superspace; 
Integrability;
}}%

\newpage

\section{Introduction}
\label{sec:introduction}

In the context of the $\AdS/\CFT$ correspondence 
\cite{Maldacena:1997re,Gubser:1998bc,Witten:1998qj}, the 
discovery of integrability is a key ingredient towards finding the exact 
spectrum of strings in $\AdS_5\times\text{S}^5$ and of composite operators in 
$\mathcal{N}=4$ SYM theory with gauge group $SU(N)$ in the planar limit, 
i.e.\ for $N\to\infty$.
As reviewed in chapters \cite{chapSpinning} and \cite{chapQstring},
on the string side of the duality
the spectrum is accessible order by order as a 
strong coupling expansion in terms of the 't Hooft coupling
by a (semi)classical analysis of string states with large quantized charges. 
It is also described in terms of respective string Bethe ans\"atze 
which are reviewed in chapter \cite{chapABA}.

In the $\mathcal{N}=4$ SYM theory, the weak coupling expansion of the 
planar spectrum, i.e.\ the conformal dimensions of composite operators, 
can be obtained by direct perturbative calculations of various 
correlation functions. The appearance of UV divergences requires
renormalization,
which then leads to a mixing among operators with the same bare conformal 
dimension. The eigenvalues of the new eigenstates under conformal rescalings
are given as the sum of the bare scaling dimension and an individual 
anomalous dimension.
The operator mixing can be extracted, e.g.\ from the 
correlation functions involving two composite operators.
Alternatively, one can directly calculate the diagrams which contribute to the
renormalization of these operators. This directly allows one to 
obtain an expression for the dilatation operator, whose
eigenvalues are the anomalous dimensions.

Perturbative calculations become very cumbersome at high loop orders
and can be avoided, if the observed integrability at one loop, 
which is reviewed in chapter \cite{chapChain}, 
also persists to higher loop orders.
The dilatation operator can then be determined, using
some very general structural information from the underlying Feynman graphs 
only and some data from the gauge Bethe ans\"atze. 
The details of this approach
are reviewed in chapter \cite{chapLR}.
%
Direct Feynman graph calculations of the dilatation operator 
in the flavour $SU(2)$ subsector to three loops and of some of its eigenvalues
and of parts of the Bethe ans\"atze also to higher loops 
provide important checks for the assumed integrability.

Even if integrability holds to all loop orders, the respective 
Bethe ans\"atze and planar dilatation operator 
allow us to
compute the anomalous dimensions only in the asymptotic regime.
In this regime, the loop order of the result 
is constrained to be strictly smaller than 
the length (the number of elementary fields) of the shortest composite 
operator involved.
At loop orders which are equal to or exceed this 
number, the so-called wrapping interactions 
\cite{Serban:2004jf,Beisert:2004hm} 
have to be considered. They are corrections due to the finite size of the 
composite operators and have their origin in the neglected higher 
genus contributions to the dilatation operator \cite{Sieg:2005kd}.
In the dual string theory the counterparts of the wrapping interactions 
are corrections due to the finite circumference of the closed string 
worldsheet cylinder \cite{Ambjorn:2005wa}. Their analyses 
are reviewed in chapters \cite{chapLuescher} and \cite{chapTBA}.

In this chapter we review the explicit Feynman graph calculations
in $\mathcal{N}=4$ SYM theory in the planar limit beyond one loop.
It is organized as follows:

In Section \ref{sec:opreno} we give a short summary of how composite 
operators are renormalized, and how the dilatation operator is defined in 
terms of the renormalization constants.

In Section \ref{sec:Dop} we then review the explicit calculations and 
tests of the dilatation operator with particular focus on calculations
beyond the first order in perturbation 
theory.\footnote{The one-loop results are reviewed in chapter 
\cite{chapChain}.} 
Only the flavour 
$SU(2)$ subsector will be considered, since most
higher loop calculations are performed within this 
subsector. 
As examples we recalculate in detail the respective one- and two-loop 
dilatation operator in $\mathcal{N}=1$ superfield formalism.
This approach is much more efficient than the originally used 
formalism without manifest supersymmetries, 
and it yields more direct 
relations between the dilatation operator and the underlying
Feynman graphs.  
We then display the result of a three-loop calculatoin and also 
summarize the existing checks of the magnon 
dispersion relation, of the structure of the dilatation operator and of 
some of its eigenvalues in the asymptotic regime at three and higher 
loops.

In Section \ref{sec:wrapint}, we review the perturbative calculations 
which consider the first wrapping corrections and hence yield
results beyond the asymptotic regime. The general strategy of these 
calculations will be explained. In this way, the four-loop
anomalous dimension for the length four
Konishi descendant in the flavour $SU(2)$ subsector could be determined. 
Further results for different operators and for the terms of highest
transcendentality 
are then summarized briefly.

In Section \ref{sec:concl} we give a concluding summary, and in 
two appendices
we present the explicit $\D$-algebra manipulations for the one- and two-loop 
calculation and the expressions for the relevant integrals.
%


\section{Renormalization of composite operators}

\label{sec:opreno}

The dilatation operator and anomalous dimensions can be obtained from
a perturbative calculation of the correlation functions which involve
the composite operators $\mathcal{O}_a$, where $a$ labels the different
operators. The encountered UV divergences
require a renormalization of the composite operators as
\begin{equation}\label{opren}
\mathcal{O}_{a,\text{ren}}(\phi_{i,\text{ren}})
=\mathcal{Z}_{a}{}^b(\lambda,\varepsilon)\mathcal{O}_{b,\text{bare}}(\phi_{i,\text{bare}})
\col\qquad 
\phi_{i,\text{ren}}=\mathcal{Z}_{i}^{1/2}\phi_{i,\text{bare}}
\col
\end{equation}
where in an appropriate basis 
$\mathcal{Z}=\unitmatrix+\delta\mathcal{Z}$, and the matrix
$\delta\mathcal{Z}$ is of order $\mathcal{O}(\lambda)$ in the 
renormalized coupling constant $\lambda$. It also 
depends on the regulator $\varepsilon$ and is in general
non-diagonal and thus leads to mixing between the different 
composite operators. The matrix element $\delta\mathcal{Z}_{a}{}^b$ is
given by the negative of the sum of the overall UV divergences of the
Feynman diagrams in which the vertices of the theory
lead to interactions between the elementary fields of operator 
$\mathcal{O}_b$, such that the resulting external field flavour and 
ordering coincide with the ones of the operator $\mathcal{O}_a$. 
One also has to consider contributions from 
wave function renormalization of the elementary fields $\phi_i$
the operators are composed of. Respective
factors $\mathcal{Z}_{i}^{1/2}$ are included within $\mathcal{Z}$.

$\mathcal{N}=4$ SYM theory can be regularized by supersymmetric 
dimensional reduction \cite{Siegel:1979wq} in $D=4-2\varepsilon$ 
dimensions. 
The coupling constant
$g_\YM$ is then accompanied by the 't Hooft mass $\mu$ in the combination 
$g_\YM\mu^{\varepsilon}$ to restore the mass dimension of the loop 
integrals. Thereby, $g_\YM$ is not renormalized and hence 
itself does not depend on $\mu$, such that superconformal invariance is
preserved. This was explicitly found to three loops
by computing the vanishing of the $\beta$-function in 
an $\mathcal{N}=1$ superfield formulation
\cite{Grisaru:1980jc,Grisaru:1980nk,Caswell:1980yi,Caswell:1980ru}.
The finiteness of $\mathcal{N}=4$ SYM theory was then later shown to 
all orders \cite{Mandelstam:1982cb,Brink:1982wv,Howe:1982tm,Howe:1983sr}.
A first argument 
was given in \cite{Ferrara:1974pu}.
In particular, the self-energy of the 
superfields is finite, i.e.\  $\mathcal{Z}_{i}^{1/2}$ is 
trivial.\footnote{This holds apart from gauge artefacts that are not 
relevant here.}
In the planar limit, where the coupling constant is $\lambda=g_\YM^2N$,
the dilatation operator is then extracted from the renormalization
constant of the composite operators in \eqref{opren} as
\begin{equation}\label{DinZ}
\mathcal{D}
=\mu\frac{\de}{\de\mu}\ln\mathcal{Z}(\lambda\mu^{2\varepsilon},\varepsilon)
=\lim_{\varepsilon\rightarrow0}\left[2\varepsilon\lambda
\frac{\de}{\de\lambda}\ln\mathcal{Z}(\lambda,\varepsilon)\right]
\pnt
\end{equation}
The logarithm of 
$\mathcal{Z}=\unitmatrix+\delta\mathcal{Z}$ has to be understood as
a formal series in powers of $\delta\mathcal{Z}$.
All poles of higher order in $\varepsilon$ must cancel in $\ln\mathcal{Z}$,
such that it only contains simple $\frac{1}{\varepsilon}$ poles.
In effect, the above description extracts the coefficient of the 
$\frac{1}{\varepsilon}$ pole of $\mathcal{Z}$, and at a given loop order 
$K$ multiplies it by a factor $2K$. This then yields the dilatation operator
as a power series
\begin{equation}\label{Dex}
\mathcal{D}=\sum_{k\ge1}g^{2k}\mathcal{D}_k
\col\qquad
g=\frac{\sqrt{\lambda}}{4\pi}
\col
\end{equation}
where for later convenience
we have absorbed powers of $4\pi$ into the definition
of a new coupling constant $g$.

\section{Dilatation operator in the \texorpdfstring{$SU(2)$}{SU(2)} subsector}

\label{sec:Dop}

$\mathcal{N}=4$ SYM theory contains six real scalar fields, four complex Weyl fermions and a gauge field that all transform in the adjoint representation of the gauge group $SU(N)$. In the following we denote these 
fields as components fields, since in a superspace formalism they 
appear as components of superfields. 
In order to build the $\mathcal{N}=1$ superfields, 
the real scalar component fields are complexified and combined together each 
with one fermion or with its complex conjugate
into three chiral superfields $\phi_i$, $i=1,2,3$
or respectively anti-chiral ones $\bar\phi_i$. The three field flavours are
transformed into each other by an $SU(3)$ subgroup of the 
$SU(4)$ R-symmetry group. The remaining gauge field and 
fermions are combined together into an $\mathcal{N}=1$ vector superfield $V$. 
An explicit expression of the $\mathcal{N}=4$ SYM action in terms of 
$\mathcal{N}=1$ superfields and the respective Feynman rules in
which the Wick rotation is included can be found, e.g.\ in \cite{Sieg:2010tz}.  
The superspace conventions are as in \cite{Gates:1983nr}, where 
also an introduction to the $\D$-algebra is given. The latter is required to
reduce the supergraphs, i.e. the Feynman diagrams in superspace, 
to ordinary spacetime objects that are located at a single point in 
the fermionic coordinates of superspace.

\subsection{Operator mixing in the \texorpdfstring{$SU(2)$}{SU(2)} subsector}
\label{subsec:fsubsec}

In the following, we denote the three chiral field flavours 
of $\mathcal{N}=4$ SYM theory by 
$\phi_i=(\phi,\psi,Z)$. The flavour $SU(2)$ subsector contains
operators which are composed of only two different types of these fields, e.g.\ $\phi$ and $Z$.
Their color indices are
all contracted with each other to yield a gauge invariant object.
In general, the gauge contractions form several cycles, and one obtains a 
multi-trace operator. Such an operator is a normal-ordered product
of single-trace operators, 
i.e.\ of operators each of which only contains a single cycle
of gauge contractions.

Mixing only occurs between those 
operators that have the same numbers of both types of fields $\phi$ and $Z$.
Then, it suffices to consider operators which contain a number of fields 
$\phi$ that does not exceed the number of fields $Z$, since the 
results for the remaining operators follow immediately by an
exchange of the role of the two fields.
Usually, the fields $\phi$ are denoted as impurities which appear 
between fields of type $Z$ within the traces over the gauge group. 
Furthermore, in the planar limit that we exclusively consider from 
now on,\footnote{See chapter \cite{chapObserv} for a review concerning 
effects of non-planarity.} the Feynman diagrams that alter the 
gauge trace structure of the composite operators are suppressed.
The renormalization of multi-trace operators then follows immediately from 
the one of their single-trace constituents. 
We can therefore restrict the analysis to single-trace operators.
In this case, the planar Feynman diagrams 
can only affect the ordering of the two different types of fields 
inside the single trace, but they cannot alter their multiplicities
and in particular the length $L$ of the composite operators that is defined as 
the total number of constituent fields. 
Flavour contractions cannot appear, since
the composite operators of the $SU(2)$ subsector
do not contain the complex conjugate fields $(\bar\phi,\bar\psi,\bar Z)$.
The $SU(2)$ subsector is closed 
under renormalization, at least perturbatively \cite{Minahan:2005jq}.
The operators
\begin{equation}\label{groneimpstate}
\tr\big(Z^L\big)\col\qquad\tr\big(\phi Z^{L-1}\big)
\end{equation}
which are the ground state and a state with a single impurity
are protected and do not acquire anomalous dimensions.
Operators which contain more than a single impurity $\phi$ undergo 
non-trivial mixing. 

Since the aforementioned operator mixing
only occurs within subsets of single-trace operators
that only differ by permutations of their field content, the 
renormalization constant $\mathcal{Z}$ and hence also 
the dilatation operator $\mathcal{D}$ can be expressed in terms of flavour
permutations that act on the constituent fields of these composite operators.  
The flavour permutations themselves 
can be written as products of permutations acting on 
nearest neighbour sites. 
For composite operators of fixed length $L$  they are given by
\cite{Beisert:2003tq}
\begin{equation}\label{permstrucdef}
\pthree{a_1}{\dots}{a_n}=\sum_{r=0}^{L-1}\perm_{a_1+r\;a_1+r+1}\cdots
\perm_{a_n+r\;a_n+r+1}
\end{equation}
and by the identity $\pone{}$ in flavour space 
that measures the length $L$ of the
composite operator it is applied to.
The structures consider the insertion of the Feynman subdiagrams in which 
elementary fields interact
at all possible positions within the single trace of the 
composite operator by the summation. Periodicity with period $L$ is 
thereby understood. No other insertions have to be considered here, since 
in the planar limit the interactions have to 
occur between adjacent fields.
 
The permutation structures \eqref{permstrucdef} admit a definition of the 
range of the interaction in flavour space obtained 
from their lists of arguments as
\begin{equation}\label{nneighbourint}
\kappa=\max_{a_1,\dots, a_n}-\min_{a_1,\dots, a_n}+2\pnt
\end{equation}
The range $\kappa$ and hence also the possible arguments $a_1,\dots,a_n$
of the permutation structures are subject to constraints from the underlying 
Feynman diagrams. In order to find the restrictions for those structures 
that can appear in the expression of the dilatation operator, 
we focus on Feynman diagrams in which the elementary interactions
occur in a single region that is simply connected also 
when the composite operator is removed from the diagram. 
These diagrams may have overall UV divergences that 
contribute with simple $\frac{1}{\varepsilon}$ poles
to the renormalization constant $\mathcal{Z}$ and hence according to 
\eqref{DinZ} also to the dilatation operator.
The remaining diagrams, in which 
the elementary interactions
occur in several non simply-connected regions after the removal of the 
composite operator, cannot contribute with
simple $\frac{1}{\varepsilon}$ poles. 
Their calculation is only required if one wants to determine 
$\mathcal{Z}$ itself completely, for example in order to check explicitly 
that in $\ln\mathcal{Z}$
all higher order poles in $\varepsilon$ cancel. Here, we will not 
consider them further and only focus on the diagrams that can contribute 
to the dilatation operator. The interaction range $R$ of a diagram 
of the latter type is defined as the number of 
adjacent elementary fields of the composite operator that
enter the single simply connected interaction region. 
It can only yield contributions with 
permutation structures \eqref{permstrucdef} that obey
the following conditions:
\begin{equation}\label{permbounds}
n\le K\col\qquad 
\kappa\le R\col\qquad R\le K+1
\col
\end{equation}
where $K$ denotes the number of loops inside the diagram.
The first inequality considers that each 
nearest-neighbour permutation is 
associated with at least one loop. 
The second condition ensures that the range of the interaction in flavour 
space does not exceed the interaction range $R$ of the 
Feynman diagram. In a third inequality $R$ itself is bounded from above
by the loop order,
since each interaction between nearest neighbour fields of the composite 
operator generates at least one loop. 
We denote the diagrams that saturate this bound, i.e.\ the ones with
interaction range $R=K+1$ as maximum range diagrams. 
Since the summation in \eqref{permstrucdef} runs over all insertion
points with periodicity $L$, the smallest integer entry can always be
fixed, e.g.\ to $1$ by shifting all $a_i$ by a common integer.
According to \eqref{permbounds} the biggest integer can then be 
at most $K$.
Further relations between the structures \eqref{permstrucdef} 
can be found in \cite{Beisert:2005wv}. The 
independent permutation structures which obey \eqref{permbounds}
then form a basis in which the $K$-loop dilatation operator can 
be written down.

The basis with elements \eqref{permstrucdef} is not the best choice 
in order to express the 
result of an explicit Feynman diagram calculation,
since the different flavour arrangements within a single 
Feynman diagram generate linear combinations of several permutation 
structures \eqref{permstrucdef} with fixed relative coefficients. 
If, instead, the generated combinations themselves are used as
basis elements, each Feynman diagram is associated with only one of them
\cite{Fiamberti:2007rj,Fiamberti:2008sh}.
The basis elements obtained from supergraphs are called chiral functions
and are defined as
 \begin{equation}\label{chifuncdef}
\begin{aligned}
\chi(a_1,\dots,a_n)=\pthree{a_1}{\dots}{a_n}\big|_{\perm\to\perm-\unitmatrix}
\col
\end{aligned}
\end{equation}
where $\perm\to\perm-\unitmatrix$ denotes a replacement of
all permutations in \eqref{permstrucdef} by the fixed combination of 
permutation and identity. The expansion of the resulting products
yields $\chi$ in terms of linear combinations of permutation structures.
For each $\chi$ we define the range of the interaction in flavour space
by applying the definition \eqref{nneighbourint} to its list of arguments.
The chiral functions capture the structure of the chiral and anti-chiral 
superfield lines of the underlying supergraphs. Hence, all supergraphs 
which only differ by the arrangement of the flavour-neutral vector fields
generate contributions with the same chiral function.
In particular, at loop order $K$ the chiral functions 
$\chi(a_1,\dots,a_n)$ with $n=K$ are associated each with a single 
Feynman graph since they do not contain any vector fields. 
We denote the respective graphs as chiral graphs.

Except of the identity $\chi()=\pone{}$, all chiral functions 
\eqref{chifuncdef} 
yield zero when they are applied to one of the protected
states in \eqref{groneimpstate}. 
The expression of the dilatation operator in terms of chiral functions should
hence not explicitly depend on $\chi()$. We will come back to this statement 
at the end of Section \ref{subsubsec:magnonscat}.

\subsection{One-loop dilatation operator}
\label{subsec:oneloop}
The one-loop calculation in the $SU(2)$ subsector was addressed
by Berenstein, Maldacena and Nastase in \cite{Berenstein:2002jq}.
They used component fields to compute the term involving 
the permutation structure $\pone{1}$, which permutes the flavour of two 
neighbouring fields. It is the maximum shuffling term at one loop, since
it shifts the position of the 
impurity by the maximum number of one site at this loop order. 
Its generalization to higher loops will be discussed in 
Section \ref{subsubsec:magnondisp}. The
remaining Feynman diagrams all contribute to the identity 
operation $\pone{}$ in flavour space
and were not computed explicitly. Instead, their contribution was 
reconstructed from the fact that the eigenvalue for the ground state 
in \eqref{groneimpstate} should be zero. 
Furthermore, the contributions
in which two neighbouring impurities interact with each other were neglected.

Using $\mathcal{N}=1$ superfields instead of component fields 
for the one-loop calculation,
only a single Feynman diagram contains a UV divergence and hence contributes
to the renormalization constant in \eqref{opren}. It is evaluated as
\begin{equation}\label{oneloopdiag}
\begin{aligned}
\settoheight{\eqoff}{$\times$}%
\setlength{\eqoff}{0.5\eqoff}%
\addtolength{\eqoff}{-11\unitlength}%
\raisebox{\eqoff}{%
\fmfframe(0,1)(-10,1){%
\begin{fmfchar*}(20,20)
\chione
\end{fmfchar*}}}
=+\lambda I_1\chi(1)\col
\end{aligned}
\end{equation}
where the bold horizontal line represents the composite operator
of arbitrary length $L\ge 2$, thereby omitting its $L-2$ elementary field
lines that do not participate in the local interaction. 
The $\D$-algebra 
manipulations are trivial in this case as explicitly displayed 
in Appendix \ref{app:Dalg}.
The resulting loop integral is given in Appendix \ref{app:integrals}.
The further one-loop diagram of gluon exchange is finite, and the 
one-loop wave function renormalization vanishes.
This is different from their behaviour in component formalism, where they 
have to be considered.
According to the description \eqref{DinZ}, the one-loop dilatation 
operator follows from \eqref{oneloopdiag} as
\begin{equation}\label{D1}
\mathcal{D}_1=-2\chi(1)
\pnt
\end{equation}

Including also the contributions to the trace operator in flavour space, 
which extends the result to the flavour $SO(6)$ 
subsector,\footnote{The flavor $SO(6)$ subsector is only closed to one loop.} 
the full one-loop calculation
in component fields was performed in \cite{Minahan:2002ve},
and the result was recognized as the Hamiltonian of a respective 
integrable Heisenberg spin chain.  


\subsection{Two-loop dilatation operator}
\label{subsec:twoloops}

A two-loop renormalization of composite operators in 
the $SU(2)$ subsector was performed in 
\cite{Gross:2002mh} in component formalism.
As in the one-loop case \cite{Berenstein:2002jq} only the 
diagrams which contribute to genuine flavour permutations were
explicitly calculated, and the coefficient of the identity
operation was determined by the condition of a vanishing
eigenvalue of the ground state \eqref{groneimpstate}.
Furthermore, the contributions
in which impurities interact with each other were neglected.

\begin{table}[t]
\begin{tabular}{c|c|c|c}
 & $R=1$ & $R=2$  & $R=3$ \\
\hline
$\chi()$
&
\idRone[2]{0.5}{0.5}{0.5}
%
&
\idRtwo{0.5}{0.5}{0.5}
&
\idRthree{0.5}{0.5}{0.5}
\\
\hline
$\chi(1)$ & $-$ & 
\chioneRtwo{0.5}{0.5}{0.5}
&
\chioneRthree{0.5}{0.5}{0.5}
\\
\hline
$\chi(1,2)$ & $-$ & $-$ & 
\chionetwoRthree{0}{0}{0}
\end{tabular}
\caption{Diagrams in $\mathcal{N}=1$ superfields (apart from eventual
  reflections) which can in principle contribute to the two-loop
  dilatation operator. Graphs which contain the vanishing 
one-loop self-energies are not drawn.
It turns out that all diagrams depicted in gray 
are also irrelevant. The two-loop chiral self-energy is finite, 
and the remaining range $R\ge2$ diagrams are 
irrelevant due to generalized finiteness conditions \cite{Sieg:2010tz}.}
\label{tab:2loopdiagrams}
\end{table}
The relevant diagrams for the complete 
two-loop calculation of the dilatation operator
in terms of $\mathcal{N}=1$ superfields
are given in Table \ref{tab:2loopdiagrams}.
The chiral self-energy 
is identically zero at one loop and finite at higher loops.
According to the generalized finiteness conditions 
derived in \cite{Sieg:2010tz}, all range $R\ge2$ 
diagrams, in which all vertices appear in loops are also finite.
This concerns all remaining diagrams in the first line and in the
second line the respective first diagram in the second and third
columns. The pole parts of the last two diagrams in this line 
in the third column cancel against each other 
\cite{Fiamberti:2007rj,Fiamberti:2008sh}. This 
cancellation is based
on the fact that, in order to obtain contributions with overall UV
divergences, a sufficient number of spinor derivatives $\D_\alpha$ and
$\barD_{\dot \alpha}$ has to remain inside the loops in order to
be transformed into spacetime derivatives.
This yields constraints on the $\D$-algebra manipulations that amount 
to the formulation of generalized finiteness conditions in \cite{Sieg:2010tz}.
All diagrams that are irrelevant due to these conditions are depicted in gray. 
We only have to compute the remaining diagrams and consider also 
their reflections where necessary. The substructures in the 
relevant range $R=2$ diagrams with
chiral function $\chi(1)$ combine into the one-loop chiral vertex 
correction that is explicitly given in \eqref{ccconeloop}. We then find
\begin{equation}\label{twoloopdiags}
\begin{aligned}
\settoheight{\eqoff}{$\times$}%
\setlength{\eqoff}{0.5\eqoff}%
\addtolength{\eqoff}{-12\unitlength}%
\raisebox{\eqoff}{%
\fmfframe(-0.5,2)(-10.5,2){%
\begin{fmfchar*}(20,20)
\chione
\fmfcmd{fill fullcircle scaled 8 shifted vloc(__vc2) withcolor black ;}
\fmfiv{plain,label=$\scriptstyle\textcolor{white}{1}$,l.dist=0}{vloc(__vc2)}
\end{fmfchar*}}}
=
\settoheight{\eqoff}{$\times$}%
\setlength{\eqoff}{0.5\eqoff}%
\addtolength{\eqoff}{-12\unitlength}%
\raisebox{\eqoff}{%
\fmfframe(0,2)(-10,2){%
\begin{fmfchar*}(20,20)
\chione
\fmfi{wiggly}{vm3{dir 180}..{dir -45}vm4}
\end{fmfchar*}}}
+
\settoheight{\eqoff}{$\times$}%
\setlength{\eqoff}{0.5\eqoff}%
\addtolength{\eqoff}{-12\unitlength}%
\raisebox{\eqoff}{%
\fmfframe(0,2)(-10,2){%
\begin{fmfchar*}(20,20)
\chione
\fmfi{wiggly}{vm3{dir 0}..{dir -135}vm5}
\end{fmfchar*}}}
+
\settoheight{\eqoff}{$\times$}%
\setlength{\eqoff}{0.5\eqoff}%
\addtolength{\eqoff}{-12\unitlength}%
\raisebox{\eqoff}{%
\fmfframe(-0.5,2)(-10.5,2){%
\begin{fmfchar*}(20,20)
\chione
\fmfi{wiggly}{vm4--vm5}
\end{fmfchar*}}}
=
-2\lambda^2I_2\chi(1)
\col\qquad
\settoheight{\eqoff}{$\times$}%
\setlength{\eqoff}{0.5\eqoff}%
\addtolength{\eqoff}{-11\unitlength}%
\raisebox{\eqoff}{%
\fmfframe(0,1)(-5,1){%
\begin{fmfchar*}(20,20)
\chionetwo
\end{fmfchar*}}}
=+\lambda^2I_2\chi(1,2)
\col
\end{aligned}
\end{equation} 
where we have to consider also the reflection of the last diagram which 
contributes with chiral function $\chi(2,1)$.
According to the description 
\eqref{DinZ}, the two-loop dilatation operator is then 
obtained by extracting the $\frac{1}{\varepsilon}$ pole of the sum of 
these diagrams and multiplying it by $-4$. With the
pole part of the respective integral $I_2$ given in 
\eqref{IKpoles} this then yields 
\begin{equation}\label{D2}
\mathcal{D}_2=4\chi(1)-2[\chi(1,2)+\chi(2,1)]
\pnt
\end{equation}
An explicit demonstration of the cancellation of the double poles
in $\ln\mathcal{Z}$ as mentioned after \eqref{DinZ} 
can be found in \cite{Sieg:2010tz}, where the 
one- and two-loop calculations were presented as a demonstration for 
the efficiency of the used approach.

\subsection{Three-loop dilatation operator}
\label{subsec:threeloops}

At three-loop order a calculation of the dilatation operator directly 
from Feynman graphs of $\mathcal{N}=1$ superfields was recently 
performed in \cite{Sieg:2010tz}. The result reads
\begin{equation}
\begin{aligned}\label{D3}
\mathcal{D}_3
&=-4(\chi(1,2,3)+\chi(3,2,1))+2(\chi(2,1,3)-\chi(1,3,2))-4\chi(1,3)\\
&\phantom{{}={}}
+16(\chi(1,2)+\chi(2,1))-16\chi(1)-4(\chi(1,2,1)+\chi(2,1,2))
\pnt
\end{aligned}
\end{equation}
It determines the 
planar spectrum in the $SU(2)$ subsector to three loops and 
hence goes beyond an earlier test of two eigenvalues \cite{Eden:2004ua}, which
 employs Anselmi's trick \cite{Anselmi:1998ms} to 
reduce the calculation to two loops. The three-loop results confirm the 
prediction from integrability in \cite{Beisert:2003tq}. 
Earlier checks of some of the three-loop eigenvalues are summarized in 
Section \ref{subsubsec:checks}.


\subsection{Partial tests at higher loops}
\label{subsec:higherloops}

To three-loop order and also beyond, certain parts of the respective 
Bethe ansatz and dilatation operator 
have been checked by direct Feynman 
diagram calculations. This concerns the so-called maximum shuffling terms, 
which contribute to the dispersion relation of the Bethe ansatz. Further
terms in the higher loop expressions of the dilatation operator have 
also been tested explicitly.

\subsubsection{Tests of the magnon dispersion relation}
\label{subsubsec:magnondisp}

Even if with the assumed integrability the $SU(2)$ 
dilatation operator itself has been 
determined only to the first few loop orders 
(see chapter \cite{chapLR} for a review),
the magnon dispersion relation of the Bethe ansatz is an all-order expression
and directly related to certain Feynman diagrams.
For a single magnon of momentum $p$ it is  given by \cite{Beisert:2004hm}
\begin{equation}\label{magnondisp}
E(p)=\sqrt{1+16g^2\sin^2\tfrac{p}{2}}-1
\col
\end{equation}
and it is fixed by the underlying symmetry algebra up to an 
unknown function of the coupling constant \cite{Beisert:2005tm}, 
which in the $\mathcal{N}=4$ SYM case essentially appears to be 
given by $g^2$ itself and has already 
been substituted accordingly.\footnote{The explicit Feynman diagram 
calculation in \cite{Sieg:2010tz} confirms that this is correct 
 to three loops.
It is non-trivial in the $\AdS_4/\CFT_3$ correspondence that is 
reviewed in chapter \cite{chapN6}.}

At a fixed loop order $K$ in the expansion of the above relation, the 
momentum dependence can be expressed as linear combination of the elements 
$\cos(k-1)p\,\sin^2\tfrac{p}{2}$ with $1\le k\le K$. In particular, the 
term with $k=K$ is generated by the so-called maximum shuffling 
diagrams, which include shifts of the position of a single impurity 
(which is a magnon in the spin chain notation) by 
the maximum number of $K$ neighbouring sites. The relevant diagrams
are given by
\begin{equation}\label{maxshuffling}
\begin{aligned}
\settoheight{\eqoff}{$\times$}%
\setlength{\eqoff}{0.5\eqoff}%
\addtolength{\eqoff}{-10.5\unitlength}%
\raisebox{\eqoff}{%
\fmfframe(1,1)(1,1){%
\begin{fmfchar*}(25,20)
\fmftop{v1}
\fmfbottom{v7}
\fmfforce{(0w,h)}{v1}
\fmfforce{(0w,0)}{v7}
\fmffixed{(0.2w,0)}{v1,v2}
\fmffixed{(0.2w,0)}{v2,v3}
\fmffixed{(0.2w,0)}{v3,v4}
\fmffixed{(0.2w,0)}{v4,v5}
\fmffixed{(0.2w,0)}{v5,v6}
\fmffixed{(0.2w,0)}{v7,v8}
\fmffixed{(0.2w,0)}{v8,v9}
\fmffixed{(0.2w,0)}{v9,v10}
\fmffixed{(0.2w,0)}{v10,v11}
\fmffixed{(0.2w,0)}{v11,v12}
\fmffixed{(0,whatever)}{vc1,vc5}
\fmffixed{(0,whatever)}{vc2,vc6}
\fmffixed{(0,whatever)}{vc3,vc7}
\fmffixed{(0,whatever)}{vc4,vc8}
\fmf{plain,tension=0.5,right=0.25}{v1,vc1}
\fmf{plain,tension=0.5,left=0.25}{v2,vc1}
\fmf{phantom,tension=0.5,right=0.25}{v2,vc2}
\fmf{plain,tension=0.5,left=0.25}{v3,vc2}
\fmf{phantom,tension=0.5,right=0.25}{v4,vc3}
\fmf{dots,tension=0.5,left=0.25}{v5,vc3}
\fmf{phantom,tension=0.5,right=0.25}{v5,vc4}
\fmf{plain,tension=0.5,left=0.25}{v6,vc4}
\fmf{plain,tension=0.5,left=0.25}{v7,vc5}
\fmf{phantom,tension=0.5,right=0.25}{v8,vc5}
\fmf{plain,tension=0.5,left=0.25}{v8,vc6}
\fmf{phantom,tension=0.5,right=0.25}{v9,vc6}
\fmf{dots,tension=0.5,left=0.25}{v10,vc7}
\fmf{phantom,tension=0.5,right=0.25}{v11,vc7}
\fmf{plain,tension=0.5,left=0.25}{v11,vc8}
\fmf{plain,tension=0.5,right=0.25}{v12,vc8}
\fmf{plain,tension=1.25,left=0}{vc1,vc5}
\fmf{plain,tension=1.25,left=0}{vc2,vc6}
\fmf{dots,tension=1.25,left=0}{vc3,vc7}
\fmf{plain,tension=1.25,left=0}{vc4,vc8}
\fmffreeze
\fmf{plain,tension=1,left=0}{vc5,vc2}
\fmf{dots,tension=1,left=0}{vc6,vc3}
\fmf{plain,tension=1,left=0}{vc7,vc4}
\fmf{plain,tension=0.5,right=0,width=1mm}{v7,v12}
\fmffreeze
\fmfposition
\end{fmfchar*}}}
\to
\lambda^{K}I_K\chi(1,2,\dots,K-1,K)
\end{aligned}
\end{equation}
and by its reflection.
When the sum of these two diagrams is applied to the eigenstate
of a single magnon with momentum $p$, it yields the eigenvalue
\begin{equation}\label{chiphaseshifts}
\lambda^{K}I_K\big[\chi(1,2,\dots,K)+\chi(K,\dots,2,1)\big]\to -8\lambda^{K}I_K\cos(K-1)p\,\sin^2\tfrac{p}{2}
\pnt
\end{equation}
According to the description \eqref{DinZ}, 
the $\frac{1}{\varepsilon}$ pole of this expression has to be multiplied
by $-2K$ to obtain its contribution to 
the magnon dispersion relation. A comparison with the respective term in 
the expansion of \eqref{magnondisp}, thereby taking into account the
relation \eqref{Dex} between the couplings, 
then makes a prediction for the $\frac{1}{\varepsilon}$ pole of the 
integral $I_K$ as
\begin{equation}
\res_0(\Kop\Rop(I_K))
=\frac{1}{(4\pi)^{2K}}\frac{(2K-2)!}{(K-1)!K!}\frac{1}{K}
\pnt
\end{equation}
The explicit expressions for the poles of $I_K$ for some $K$ are 
listed in \eqref{IKpoles}. They are consistent with this result.

In \cite{Gross:2002su} it was shown that at generic loop order
the pole structure of the maximum shuffling diagrams in component fields
is in accord 
with the BMN square root formula \cite{Berenstein:2002jq}.
The latter was proposed as an all-order expression for the anomalous dimensions
in the so-called BMN limit, where the length $L$ of the operators and the coupling $g$ become infinite $L,g\to\infty$, thereby keeping fixed the numbers of 
impurities inside the operators and also the effective coupling constant 
$g'=\frac{g}{L}$.
For magnon momenta $p_j=\frac{2\pi n_j}{L}\ll 1$
the dispersion relation \eqref{magnondisp} yields the individual
contributions of each magnon $j$ with mode number $n_j$ 
to the BMN square root formula. 
Since the scattering of magnons is neglected, 
their momenta $p_j$ assume a simple form and  
are solutions of the originally proposed Bethe 
equations \cite{Beisert:2004hm} with a magnon S-matrix that becomes trivial 
in the BMN limit. 
However, these Bethe equations do not yield the anomalous 
dimensions of $\mathcal{N}=4$ SYM theory since the S-matrix is incomplete.
One has to consider the 
so-called dressing phase \cite{Beisert:2006ez} that first
appeared at strong coupling \cite{Arutyunov:2004vx,Hernandez:2006tk} but 
is important also at weak coupling \cite{Beisert:2006ez,Bern:2006ew}, 
where it alters the magnon momenta
at order $\mathcal{O}(g^6)$.\footnote{The dressing phase is reviewed in 
chapter \cite{chapSProp}.} 
Due to the dressing phase, the S-matrix violates perturbative 
BMN scaling, i.e.\ its perturbative expansion diverges if after 
the replacement $g\to g'L$ the limit $L\to\infty$ is taken, 
thereby keeping $g'$ fixed and small.
The Bethe equations involving this S-matrix then yield anomalous dimensions
that violate perturbative BMN scaling from four loops on.
However, the BMN square root formula obeys this scaling, and hence it cannot
describe the anomalous dimensions of operators with two or more impurities
beyond three loops.\footnote{This breakdown is independent of the general restriction of the Bethe ansatz to the asymptotic regime that requires a 
termination of the expansion at a loop order $K\le L-1$ to avoid the wrapping corrections.}
Since the dressing phase only affects the scattering of magnons, 
all tests and derivations of the BMN square root formula 
that rely on the calculation of phase shifts of a single magnon are insensitive
to this failure and succeed. This concerns the previously mentioned 
all-order test of the maximum shuffling terms \cite{Gross:2002su} 
and also an all order derivation 
employing the $\mathcal{N}=1$ superfield formalism
\cite{Santambrogio:2002sb}.
It would be more appropriate 
to say that in these calculations the magnon dispersion relation in the 
BMN limit is obtained.

The magnon dispersion relation \eqref{magnondisp} describes 
the free propagation of one magnon. It it thus built up from 
all Feynman diagrams with chiral functions that do not yield 
a vanishing result when applied to the single magnon momentum eigenstate.
The number of impurities of the composite operator
sets an upper bound on the number of bubbles formed by two 
neighbouring lines of the composite operator inside the Feynman diagrams. 
Such a bubble appears for example 
in the lower right corner of the graph in \eqref{maxshuffling}, and 
it vanishes unless the two involved 
field flavours are different. The diagrams contributing to the magnon 
dispersion relation hence must not contain more than
one of these bubbles. 
This restricts their chiral functions to 
$\chi(1,\dots,k)$ and $\chi(k,\dots,1)$ after the identities for 
the permutation structures \eqref{permstrucdef} found in
\cite{Beisert:2005wv} have been used to simplify the chiral functions, e.g.\ as
$\chi(1,2,1)=\chi(2,1,2)=\chi(1)$ in the three loop result \eqref{D3}.
All-order expressions for the coefficients of these terms in the 
dilatation operator then follow directly from the magnon dispersion
relation \eqref{magnondisp} and can be found in \cite{Sieg:2010tz}.
It should be stressed that the aforementioned contributions 
also yield non-vanishing results when additional magnons are present
outside of the $k+1$ interacting legs. They therefore also contribute to
the magnon S-matrix.

\subsubsection{Tests of magnon scattering}
\label{subsubsec:magnonscat}

The Feynman diagrams that vanish for a single magnon state, but are 
non-vanishing if two or more magnons are present within their respective
interaction ranges, should exclusively be associated with the magnon S-matrix.
Their contributions appear together with the ones of the 
aforementioned maximum and non-maximum
shuffling terms in the dilatation operator. 
In the $SU(2)$ subsector they first show up at 
three-loops as the contribution with chiral function $\chi(1,3)$ in 
\eqref{D3}.\footnote{A two-loop test of the S-matrix of the $SL(2)$
subsector can be found in \cite{Eden:2005bt}.}
The further chiral functions $\chi(2,1,3)$, $\chi(1,3,2)$ are also 
associated with magnon scattering, but they only appear in a 
combination that is associated with a similarity transformation, 
i.e.\ a change in the basis of operators 
\cite{Beisert:2003tq,Beisert:2005wv}, that does not affect the eigenvalues.

As a more complicated example, we consider the four-loop dilatation operator.
It can be determined from the underlying integrability as 
reviewed in chapter \cite{chapLR}. 
In the basis of the chiral functions \eqref{chifuncdef} it reads
\begin{equation}
\begin{aligned}\label{D4}
\mathcal{D}_4&={}+{}200\chi(1)
-150[\chi(1,2)+\chi(2,1)]
+8(10+\epsilon_{3a})\chi(1,3)
-4\chi(1,4)\\
&\phantom{{}={}}
+60[\chi(1,2,3)+\chi(3,2,1)]\\
&\phantom{{}={}}
+(8+2\beta+4\epsilon_{3a}-4i\epsilon_{3b}+2i\epsilon_{3c}-4i\epsilon_{3d})
\chi(1,3,2)\\
&\phantom{{}={}}
+(8+2\beta+4\epsilon_{3a}+4i\epsilon_{3b}-2i\epsilon_{3c}+4i\epsilon_{3d})
\chi(2,1,3)\\
&\phantom{{}={}}
-(4+4i\epsilon_{3b}+2i\epsilon_{3c})[\chi(1,2,4)+\chi(1,4,3)]\\
&\phantom{{}={}}
-(4-4i\epsilon_{3b}-2i\epsilon_{3c})[\chi(1,3,4)+\chi(2,1,4)]\\
&\phantom{{}={}}-(12+2\beta+4\epsilon_{3a})\chi(2,1,3,2)\\
&\phantom{{}={}}
+(18+4\epsilon_{3a})[\chi(1,3,2,4)+\chi(2,1,4,3)]\\
&\phantom{{}={}}
-(8+2\epsilon_{3a}+2i\epsilon_{3b})[\chi(1,2,4,3)+\chi(1,4,3,2)]\\
&\phantom{{}={}}
-(8+2\epsilon_{3a}-2i\epsilon_{3b})[\chi(2,1,3,4)+\chi(3,2,1,4)]\\
&\phantom{{}={}}
-10[\chi(1,2,3,4)+\chi(4,3,2,1)]
\pnt
\end{aligned}
\end{equation}
The coefficients $\epsilon_i$, $i=3a,3b,3c,3d$ in the above result 
are not fixed by the construction and parameterize the previously mentioned
similarity transformations. The coefficient $\beta$ is the leading 
term of the previously mentioned dressing phase. 
The magnon dispersion relation is encoded in 
the first two terms in the first line, the 
second line and the last line.
The further contributions should be associated with magnon scattering.
As the contributions from the maximum shuffling diagrams \eqref{maxshuffling}
in the last line, also the other terms in the last four lines have chiral 
functions that saturate all the bounds in \eqref{permbounds}. 
Hence, the underlying
Feynman diagrams are chiral and of maximum range and their contributions 
can be calculated as easily as the one of the maximum 
shuffling terms \eqref{maxshuffling}.

The term in \eqref{D4} with chiral function $\chi(2,1,3,2)$
only satisfies the first bound in 
\eqref{permbounds}, i.e.\ the underlying Feynman diagram is chiral 
but it is not of maximum range.
It involves the leading coefficient $\beta$ of the 
dressing phase, which can be determined from 
an evaluation of the respective diagram
\begin{equation}
\begin{aligned}
\settoheight{\eqoff}{$\times$}%
\setlength{\eqoff}{0.5\eqoff}%
\addtolength{\eqoff}{-11\unitlength}%
\raisebox{\eqoff}{%
\fmfframe(-1,1)(-1,1){%
\begin{fmfchar*}(20,20)
\fmftop{v1}
\fmfbottom{v5}
\fmfforce{(0.125w,h)}{v1}
\fmfforce{(0.125w,0)}{v5}
\fmffixed{(0.25w,0)}{v1,v2}
\fmffixed{(0.25w,0)}{v2,v3}
\fmffixed{(0.25w,0)}{v3,v4}
\fmffixed{(0.25w,0)}{v5,v6}
\fmffixed{(0.25w,0)}{v6,v7}
\fmffixed{(0.25w,0)}{v7,v8}
\fmffixed{(0,whatever)}{vc1,vc5}
\fmffixed{(0,whatever)}{vc2,vc3}
\fmffixed{(0,whatever)}{vc3,vc6}
\fmffixed{(0,whatever)}{vc6,vc7}
\fmffixed{(0,whatever)}{vc4,vc8}
\fmffixed{(0.5w,0)}{vc1,vc4}
\fmffixed{(0.5w,0)}{vc5,vc8}
\fmf{plain,tension=1,right=0.125}{v1,vc1}
\fmf{plain,tension=0.25,right=0.25}{v2,vc2}
\fmf{plain,tension=0.25,left=0.25}{v3,vc2}
\fmf{plain,tension=1,left=0.125}{v4,vc4}
\fmf{plain,tension=1,left=0.125}{v5,vc5}
\fmf{plain,tension=0.25,left=0.25}{v6,vc6}
\fmf{plain,tension=0.25,right=0.25}{v7,vc6}
\fmf{plain,tension=1,right=0.125}{v8,vc8}
  \fmf{plain,tension=0.5}{vc1,vc3}
  \fmf{plain,tension=0.5}{vc2,vc3}
  \fmf{plain,tension=0.5}{vc3,vc4}
  \fmf{plain,tension=0.5}{vc5,vc7}
  \fmf{plain,tension=0.5}{vc6,vc7}
  \fmf{plain,tension=0.5}{vc7,vc8}
  \fmf{plain,tension=2}{vc1,vc5}
  \fmf{plain,tension=2}{vc4,vc8}
  \fmf{phantom,tension=2}{vc5,vc4}
\fmffreeze
\fmfposition
\fmf{plain,tension=1,left=0,width=1mm}{v5,v8}
\fmffreeze
\end{fmfchar*}}}
\to\lambda^4I_\beta\chi(2,1,3,2)
\end{aligned}
\end{equation}
if the coefficient $\epsilon_{3a}$ of 
the similarity transformations is known. One finds $\epsilon_{3a}=-4$ 
for example by computing the diagram which 
generates $\chi(1,3,2,4)$ or $\chi(2,1,4,3)$. 
With the pole part of the integral $I_\beta$ given in \eqref{Ipoles},
the leading coefficient of the dressing phase is then
determined as $\beta=4\zeta(3)$. The result was obtained in
\cite{Beisert:2007hz}, using component formalism. It agrees with one
of the proposals in \cite{Beisert:2006ez} and with the result extracted from a
four-loop calculation of a four-point amplitude in \cite{Bern:2006ew}.

It is also relatively easy to compute the terms with chiral 
functions which only saturate the second and third
bound in \eqref{permbounds},
i.e.\ all terms in \eqref{D4} with chiral functions 
that contain $1$ and $4$ in their lists of arguments and
hence only stem from Feynman diagrams of maximum range
$R=5$. This calculation was performed in 
\cite{Fiamberti:2007rj,Fiamberti:2008sh} in $\mathcal{N}=1$ 
superfield formalism in the context of calculating the first wrapping
correction to be discussed below. 
The results yield an overdetermined
system of equations that uniquely fixes the coefficients $\epsilon_i$ 
and provides non-trivial checks of the remaining coefficients 
that are fixed by the underlying integrability. 
The analogous calculation of the $R=6$ diagrams 
at five loops can be found in \cite{Fiamberti:2009jw}.

The expressions \eqref{D1}, \eqref{D2}, \eqref{D3} and \eqref{D4} 
do not depend on the identity $\chi()$. This guarantees that the 
anomalous dimension of the BPS operators \eqref{groneimpstate}
are zero. The generalized finiteness conditions in \cite{Sieg:2010tz}
predict this to all orders and relate it to the finiteness
of the chiral self-energy, i.e.\ to the preservation of conformal invariance.

\subsubsection{Checks of eigenvalues}
\label{subsubsec:checks}

To three loops the results \eqref{D1}, \eqref{D2} and \eqref{D3} 
for the dilatation operator have been obtained by direct Feynman diagram 
calculations. At higher loops, only the terms that saturate at least 
one of the bounds in \eqref{permbounds} have been
tested as described above. 
Further checks concern the eigenvalues of the dilatation operator for 
some composite operators. They should match with the anomalous dimensions 
obtained in direct Feynman diagram calculations.

Of particular interest is thereby the Konishi 
supermultiplet. As superconformal primary it contains 
the $\mathcal{N}=1$ Konishi operator \cite{Konishi:1983hf} 
that has bare scaling 
dimension $\Delta_0=2$ and reads
\begin{equation}\label{Konishiop}
\mathcal{K}=\tr\big(\e^{-g_\YM V}\bar\phi_i\e^{g_\YM V}\phi^i\big)
\pnt
\end{equation}
This operator is not chiral, and hence all its superfield 
components lie beyond the $SU(2)$ subsector. However, the Konishi 
supermultiplet also contains an operator of this subsector.
In order to find it, one has to select the
level four descendant of bare dimension $\Delta_0=4$ that is chiral and
pick out the relevant $SU(4)$ R-symmetry component given by
\begin{equation}\label{Kdesc}
\tr\big(\comm{\phi}{Z}\comm{\phi}{Z}\big)
\pnt
\end{equation}
It contains as lowest superfield component the respective operator built out of
the two scalar fields of the flavour $SU(2)$ subsector.

All members of a superconformal multiplet acquire the same anomalous 
dimension. For the Konishi multiplet it is given 
to four loops in \eqref{gammaK}. 
The one- and two-loop contributions were 
obtained by explicit 
Feynman diagram calculations in \cite{Anselmi:1996mq,Anselmi:1996dd} and
\cite{Bianchi:1999ge,Bianchi:2000hn}, and then also by an OPE analysis in
\cite{Arutyunov:2001mh}, see also \cite{Bianchi:2001cm}.
These results are also found for a twist-two operator with conformal 
spin $S=2$ that appears within another level four descendant of the 
Konishi multiplet. It belongs to the closed $SL(2)$ subsector that 
contains certain operators with general twist and 
conformal spin $S$. 
For twist-two operators with generic $S$, the result to two loops 
has been obtained from Feynman diagrams in \cite{Kotikov:2003fb}.
At three loops it could be extracted \cite{Kotikov:2004er} 
as the terms with highest transcendentality,
i.e.\ with highest degrees of the harmonic sums, 
from the NNLO QCD result for the non-singlet splitting functions of QCD 
\cite{Moch:2004pa}.
The truncation of the QCD result is based on the observation 
\cite{Kotikov:2002ab} that due to 
special properties of the DGLAP and BFKL equations in 
$\mathcal{N}=4$ SYM theory a mixing between functions of 
different transcendentality degrees does not occur.
Specializing to $S=2$, the extracted result agrees with the three-loop 
contribution in \eqref{gammaK}. 
When the dilatation operator given in \eqref{D1}, \eqref{D2} and \eqref{D3}
is applied to the state \eqref{Kdesc}, it also 
correctly yields the result in \eqref{gammaK}.\footnote{At four and higher loops this is no longer the case since the 
wrapping interactions have to be considered. 
This will be discussed in Section \ref{sec:wrapint}.}
In fact, the three-loop term was first predicted in 
\cite{Beisert:2003tq}, where the dilatation operator was constructed from 
integrability. Later, an explicit Feynman diagram calculation 
\cite{Eden:2004ua}, which employs Anselmi's trick \cite{Anselmi:1998ms} 
to reduce the calculation to two loops, led to the same result. 
The calculation in \cite{Sieg:2010tz} also confirms the result and furthermore
fixes the planar three-loop spectrum of all composite single-trace operators 
of the flavour $SU(2)$ subsector from field theory by a direct
Feynman diagram calculation of the dilatation operator.

The previously mentioned twist-two operators of the $SL(2)$ sector 
are very important for tests of 
the $\AdS/\text{CFT}$ correspondence and the underlying integrability. 
These tests are reviewed in chapter \cite{chapTwist}.
In particular, the results in the strict $S\to\infty$ limit are not 
modified by wrapping interactions.
At finite $S$ such modifications occur. The simplest example is $S=2$, i.e.\
the operator which appears in the Konishi multiplet.
Its anomalous dimension is affected by wrapping interactions at four loops and 
beyond. 

\section{Wrapping interactions}

\label{sec:wrapint}

In the following we briefly summarize the calculations of the previously 
mentioned wrapping interactions. A more detailed review 
is given by \cite{Fiamberti:2010fw}.

\begin{figure}[h]
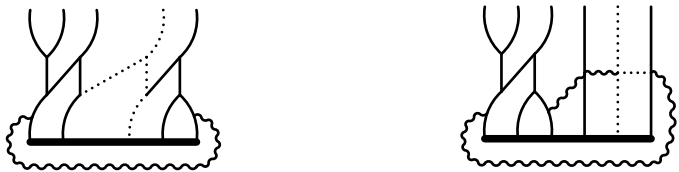

\vspace{-0.25\baselineskip}
\begin{center}
\settoheight{\eqoff}{$\times$}%
\setlength{\eqoff}{0.5\eqoff}%
\addtolength{\eqoff}{-13\unitlength}%
\raisebox{\eqoff}{%
\fmfframe(4,1)(3.5,5){%
\begin{fmfchar*}(25,20)
\fmftop{v1}
\fmfbottom{v7}
\fmfforce{(0w,h)}{v1}
\fmfforce{(0w,0)}{v7}
\fmffixed{(0.2w,0)}{v1,v2}
\fmffixed{(0.2w,0)}{v2,v3}
\fmffixed{(0.2w,0)}{v3,v4}
\fmffixed{(0.2w,0)}{v4,v5}
\fmffixed{(0.2w,0)}{v5,v6}
\fmffixed{(0.2w,0)}{v7,v8}
\fmffixed{(0.2w,0)}{v8,v9}
\fmffixed{(0.2w,0)}{v9,v10}
\fmffixed{(0.2w,0)}{v10,v11}
\fmffixed{(0.2w,0)}{v11,v12}
\fmffixed{(0,whatever)}{vc1,vc5}
\fmffixed{(0,whatever)}{vc2,vc6}
\fmffixed{(0,whatever)}{vc3,vc7}
\fmffixed{(0,whatever)}{vc4,vc8}
\fmf{plain,tension=0.5,right=0.25}{v1,vc1}
\fmf{plain,tension=0.5,left=0.25}{v2,vc1}
\fmf{phantom,tension=0.5,right=0.25}{v2,vc2}
\fmf{plain,tension=0.5,left=0.25}{v3,vc2}
\fmf{phantom,tension=0.5,right=0.25}{v4,vc3}
\fmf{dots,tension=0.5,left=0.25}{v5,vc3}
\fmf{phantom,tension=0.5,right=0.25}{v5,vc4}
\fmf{plain,tension=0.5,left=0.25}{v6,vc4}
\fmf{plain,tension=0.5,left=0.25}{v7,vc5}
\fmf{phantom,tension=0.5,right=0.25}{v8,vc5}
\fmf{plain,tension=0.5,left=0.25}{v8,vc6}
\fmf{phantom,tension=0.5,right=0.25}{v9,vc6}
\fmf{dots,tension=0.5,left=0.25}{v10,vc7}
\fmf{phantom,tension=0.5,right=0.25}{v11,vc7}
\fmf{plain,tension=0.5,left=0.25}{v11,vc8}
\fmf{plain,tension=0.5,right=0.25}{v12,vc8}
\fmf{plain,tension=1.25,left=0}{vc1,vc5}
\fmf{plain,tension=1.25,left=0}{vc2,vc6}
\fmf{dots,tension=1.25,left=0}{vc3,vc7}
\fmf{plain,tension=1.25,left=0}{vc4,vc8}
\fmffreeze
\fmf{plain,tension=1,left=0}{vc5,vc2}
\fmf{dots,tension=1,left=0}{vc6,vc3}
\fmf{plain,tension=1,left=0}{vc7,vc4}
\fmf{plain,tension=0.5,right=0,width=1mm}{v7,v12}
\fmffreeze
\fmfposition
\fmfipath{p[]}
\fmfiset{p1}{vpath(__v7,__vc5)}
\fmfiset{p2}{vpath(__v12,__vc8)}
\fmfipair{wz[]}
\fmfiequ{wz2}{point length(p2)/2 of p2}
\svertex{wz1}{p1}
\svertex{wz2}{p2}
\wigglywrap{wz1}{v7}{v12}{wz2}
\fmfposition
\end{fmfchar*}}}
\hspace{3cm}
\settoheight{\eqoff}{$\times$}%
\setlength{\eqoff}{0.5\eqoff}%
\addtolength{\eqoff}{-12.5\unitlength}%
\settoheight{\eqofftwo}{$\times$}%
\setlength{\eqofftwo}{0.5\eqofftwo}%
\addtolength{\eqofftwo}{-13\unitlength}%
\raisebox{\eqoff}{%
\fmfframe(4,1)(3.5,5){%
\begin{fmfchar*}(25,20)
\fmftop{v1}
\fmfbottom{v7}
\fmfforce{(0w,h)}{v1}
\fmfforce{(0w,0)}{v7}
\fmffixed{(0.2w,0)}{v1,v2}
\fmffixed{(0.2w,0)}{v2,v3}
\fmffixed{(0.2w,0)}{v3,v4}
\fmffixed{(0.2w,0)}{v4,v5}
\fmffixed{(0.2w,0)}{v5,v6}
\fmffixed{(0.2w,0)}{v7,v8}
\fmffixed{(0.2w,0)}{v8,v9}
\fmffixed{(0.2w,0)}{v9,v10}
\fmffixed{(0.2w,0)}{v10,v11}
\fmffixed{(0.2w,0)}{v11,v12}
\fmffixed{(0,whatever)}{vc1,vc3}
\fmffixed{(0,whatever)}{vc2,vc4}
\fmf{plain,tension=0.5,right=0.25}{v1,vc1}
\fmf{plain,tension=0.5,left=0.25}{v2,vc1}
\fmf{phantom,tension=0.5,right=0.25}{v2,vc2}
\fmf{plain,tension=0.5,left=0.25}{v3,vc2}
\fmf{plain,tension=0.5,left=0.25}{v7,vc3}
\fmf{phantom,tension=0.5,right=0.25}{v8,vc3}
\fmf{plain,tension=0.5,left=0.25}{v8,vc4}
\fmf{plain,tension=0.5,right=0.25}{v9,vc4}
\fmf{plain,tension=1.25,left=0}{vc1,vc3}
\fmf{plain,tension=1.25,left=0}{vc2,vc4}
\fmffreeze
\fmf{plain,tension=1,left=0}{vc2,vc3}
\fmf{plain}{v4,v10}
\fmf{dots}{v5,v11}
\fmf{plain}{v6,v12}
\fmf{plain,tension=0.5,right=0,width=1mm}{v7,v12}
\fmffreeze
\fmfposition
\fmfipath{p[]}
\fmfipair{wz[]}
\fmfiset{p9}{vpath(__v9,__vc4)}
\fmfiset{p4}{vpath(__v4,__v10)}
\fmfiset{p5}{vpath(__v5,__v11)}
\fmfiset{p6}{vpath(__v6,__v12)}
\fmfiset{p7}{vpath(__v7,__vc3)}
\svertex{wz9}{p9}
\svertex{wz4}{p4}
\svertex{wz5}{p5}
\svertex{wz6}{p6}
\svertex{wz7}{p7}
\fmfi{wiggly}{wz4..wz5}
\fmfi{dots}{wz5..wz6}
\fmfi{wiggly}{wz9..wz4}
\wigglywrap{wz7}{v7}{v12}{wz6}
\end{fmfchar*}}}
\caption{Wrapping diagrams that generate contributions of highest transcendentality at leading wrapping order.\label{fig:hightranswrap}}
\end{center}
\vspace{-1.25\baselineskip}
\end{figure}

The Bethe ans\"atze or the dilatation operator yield reliable results
for the anomalous dimensions in the asymptotic limit
only. The origin and precise form of this restriction can be understood 
by recalling the construction from Feynman diagrams. In Section \ref{sec:Dop}
it was argued that at a given loop order $K$ the dilatation operator is 
determined from Feynman diagrams with range $R\le K+1$, which lead to 
flavour permutations with range $\kappa\le R$. 
For the construction of the diagrams, it is thereby implicitly assumed 
that the length $L$ of the involved composite operators
is at least as big as the maximal interaction range $K+1$.
Therefore, an application of the dilatation operator 
to composite operators of length $L$
can in general only yield the correct anomalous dimensions
in the asymptotic limit, i.e.\ to a loop order $K\le L-1$.
At $K\ge L$ loops, the assumption of a sufficient length of the
involved composite operators becomes invalid, and therefore
contributions from diagrams with interaction range $R>L$
should be removed from the dilatation operator. 
Instead, there are contributions from new diagrams that are built with the 
operators of the respective lower length $L$.
The new diagrams are called wrapping diagrams since, due to the insufficient 
length of the composite operators, the interactions wrap around them.
Two examples of such diagrams are depicted in Figure
\ref{fig:hightranswrap}.
Beyond the asymptotic limit, the dilatation operator 
explicitly depends on the length $L$ of the 
composite operators it is applied to. More precisely,
the coefficients of the chiral functions 
in the expression of the dilatation operator become functions of $L$ at loop 
orders $K\ge L$, while in the asymptotic limit
they are constants, and the dilatation operator
depends on the length only via the permutation 
structures \eqref{permstrucdef}.

The appearance of wrapping interactions is closely connected to the 
truncation of the genus $h$ expansion of the dilatation operator beyond 
the planar $h=0$ contribution \cite{Sieg:2005kd}.
If in a planar wrapping diagram the composite operator is replaced by a 
longer operator, the additional 
fields lines cannot leave the diagram without crossing any other lines, i.e.\
it becomes a diagram of genus $h=1$. The appearing wrapping diagrams
hence come from certain genus $h=1$ contributions to the dilatation 
operator, which become planar when it is applied to a sufficiently 
short composite operator.
Wrapping diagrams appear at all orders in the genus expansion
of the dilatation operator. They are of genus $h+1$ in the asymptotic regime
and encode the finite size effects at genus $h$. 
The planar wrapping diagrams are special since they can 
be projected out of all genus one contributions by introducing  
spectator fields \cite{Sieg:2005kd}.
While in general for higher genus diagrams the notion of the range of
the interaction is not meaningful, it is still well defined for the 
subset of genus one diagrams when they become the planar wrapping diagrams.
Integrability seems to persist, even if in general at higher genus
its breakdown is expected 
\cite{Beisert:2003tq}.\footnote{In chapter \cite{chapObserv} the analyses of 
higher genus contributions are reviewed.}

In order to obtain the anomalous dimensions beyond the asymptotic regime,
one should not abandon the dilatation operator as obtained from the 
underlying integrability at loop orders $K\ge 4$ and compute all Feynman 
diagrams. Instead, 
the considerations at the beginning of this section imply 
that the dilatation operator is still useful, 
since it can be corrected for an
application to composite operators of shorter length $L$.
First, at each loop order $K$ all contributions from Feynman graphs of longer 
range $K+1\ge R> L$ have to be removed. 
Then, contributions from 
the wrapping interactions have to be added. 

This procedure is 
particularly powerful at the critical order $K=L$ where wrapping arises for 
the first time, since only relatively few Feynman diagrams of restricted 
topology have to be computed explicitly. Most diagrams are captured
automatically by those terms in the dilatation operator that are not 
removed in the modification process. Also, the only contributions 
that one has to remove from the dilatation operator are the ones
that come from Feynman diagrams with maximum range $R=K+1$. 
It is convenient to divide these diagrams
according to their range of interaction 
in flavour space $\kappa$ into two classes. 
The first class contains diagrams with $\kappa=R=K+1$, i.e.\
according to the definition of $\kappa$ in \eqref{nneighbourint}  
their range $R$ is encoded within the list of arguments of their 
chiral functions. The second class collects all the remaining diagrams
with $\kappa<R=K+1$.  Such Feynman diagrams contain a chiral structure 
with interaction range 
$\kappa$, and the remaining $R-\kappa$ neighbouring field lines are 
connected with it and with each other only by vector fields. 
Since the latter are flavour neutral, the range $R$ of these diagrams is not
captured by the chiral functions. It was shown in \cite{Fiamberti:2008sh} 
in the $\mathcal{N}=1$ superfield formalism that the 
diagrams of the second class do 
not contribute to the dilatation operator: either they are finite or 
their overall UV divergences cancel against each other. 
This is also an implication of the generalized finiteness conditions
derived in \cite{Sieg:2010tz}. In Section \ref{subsec:twoloops} we have 
already used the results when we disregarded the two-loop diagrams 
with $R=3$ but $\kappa<3$
in the first two rows of the last column of Table \ref{tab:2loopdiagrams}. 
The diagrams of the first class
that have $\kappa=R=K+1$ are the only 
maximum range diagrams that contribute with their overall UV divergences. 
These contributions can be easily identified and removed from 
the expression of the dilatation operator, 
since their chiral functions are of maximum range. The subtraction 
procedure becomes almost trivial: one just has to remove all contributions 
with chiral functions that have $1$ and $K$ within their list of arguments. 
This does not require the calculation of any Feynman diagrams.
For example, in the four-loop expression \eqref{D4} one removes
the last contribution in the first line and the ones in the fifth, sixth
and the last four lines.
The eigenvalues of the subtracted dilatation operator are no 
longer independent of the scheme coefficients $\epsilon_i$, which have to 
be fixed by calculating at least some of the diagrams with range 
$R=K+1$.
If one could compute the wrapping interactions that have to be added
to the subtracted dilatation operator also as functions of $\epsilon_i$, 
the eigenvalues of the resulting operator should not depend on the $\epsilon_i$.
However, the calculation of the wrapping interactions takes place in a 
scheme fixed by the use of $\mathcal{N}=1$ supergraphs, and therefore
the $\epsilon_i$ in the subtracted dilatation operator have to assume 
the respective values. 
Finally, it is important to remark that the 
simplicity of the
subtraction procedure is only guaranteed if
chiral functions \eqref{chifuncdef} are used
as basis elements.
If, instead, the basis of permutation structures \eqref{permstrucdef} 
is used, the subtraction of the contribution from a Feynman diagram with 
$R=K+1$ affects the coefficients of several permutation structures
also with different flavour interaction ranges $\kappa\le R$
in the dilatation 
operator.\footnote{In the context of the BMN matrix model a 
subtraction attempt
was made in \cite{Fischbacher:2004iu}. It does not lead to the correct 
result, since the necessary modifications of the contributions with 
permutation structures of lower range and the addition of the wrapping 
diagrams was not performed.}

The aforementioned method was first introduced and used in 
\cite{Fiamberti:2007rj}, with the details given in \cite{Fiamberti:2008sh}, 
in the case $K=L=4$, i.e.\ for the four-loop anomalous dimension of the 
Konishi operator. In $\mathcal{N}=4$ SYM theory it is the simplest case 
where wrapping arises. The calculation starts from the four-loop 
asymptotic dilatation operator \eqref{D4} and modifies it for an application 
to the length four Konishi descendant of the flavour $SU(2)$ subsector 
\eqref{Kdesc} in order to determine the correct eigenvalue 
\cite{Fiamberti:2007rj,Fiamberti:2008sh}. 
Including also the lower orders,
the anomalous dimension of the Konishi operator to four-loops was then 
determined as
\begin{equation}
\label{gammaK}
\gamma=12g^2-48g^4+336g^6+(-2496+576\zeta(3)-1440\zeta(5))g^8
\col
\end{equation}
where the full conformal dimension is obtained as
$\Delta=\Delta_0+\gamma$ with the bare scaling dimension $\Delta_0$
as described in Section \ref{subsubsec:checks}. 
The four-loop contribution has also been obtained from
a generalized L\"uscher formula \cite{Bajnok:2008bm}. This approach is reviewed
in chapter \cite{chapLuescher}.
Furthermore, it was later also found in a computer-based calculation 
in component formalism \cite{Velizhanin:2008jd}.
The matching of the Feynman diagram and L\"uscher based calculations 
provides the first test of $\AdS/\CFT$ and the underlying integrability 
beyond the 
asymptotic limit. It is also reproduced by the recently proposed 
$Y$-system \cite{Gromov:2009bc,Gromov:2009tv}, which is derived from the 
thermodynamic Bethe ansatz (TBA)
\cite{Bombardelli:2009ns,Arutyunov:2007tc,Arutyunov:2009zu} and is
a candidate to capture the full planar spectrum of $\mathcal{N}=4$ SYM theory.
The TBA and Y-system  are reviewed, respectively, in chapters
\cite{chapTBA} and \cite{chapTrans}.
Earlier attempts to describe the wrapping effects 
in terms of integrable systems are included in chapter \cite{chapLR}.


In \cite{Bajnok:2008qj} the result \eqref{gammaK} which also holds for 
the earlier mentioned twist-two operator with conformal spin $S=2$ 
has been generalized to arbitrary $S$. When analytically continued to $S=-1$, 
it  yields the correct pole structure as predicted from the BFKL 
equation. 

A result for the five-loop anomalous dimension of the Konishi
operator has been obtained in impressive calculations on the basis of the 
generalized L\"uscher formula \cite{Bajnok:2009vm} and the 
TBA \cite{Arutyunov:2010gb,Balog:2010xa}. 
Also this result has been generalized to arbitrary spin $S$, and it
is in accord with the pole structure from the BFKL equation 
\cite{Lukowski:2009ce}.
To obtain the five-loop result for the Konishi multiplet
from a Feynman diagram calculation is very difficult, even with the 
universal cancellation mechanisms discovered in \cite{Sieg:2010tz}.
Instead, a five-loop result for the $L=5$ operator 
$\tr\big(\comm{\phi}{Z}\comm{\phi}{Z}Z\big)$
which is in the same supermultiplet as certain twist-three operators
has been computed \cite{Fiamberti:2009jw},
and it agrees with the result from the generalized L\"uscher formula
\cite{Beccaria:2009eq}. 
The six-loop results for the twist-three
operators with generic conformal spin $S$ has recently become available
\cite{Velizhanin:2010cm}.

Beyond the asymptotic limit, the contributions of 
highest transcendentality, i.e.\ which contain the $\zeta$-function with 
biggest argument, are generated entirely by the wrapping interactions.
In the four-loop result in \eqref{gammaK} this is the term with $\zeta(5)$. 
Its generalization to twist-two operators with generic 
conformal spin $S$ has been obtained from a 
Feynman diagram calculation in component formalism in 
\cite{Velizhanin:2008pc}.
At generic loop and critical wrapping order $K=L$ 
the highest transcendentality degree of the wrapping diagrams 
is $2K-3$ compared to $2K-5$ of the dressing phase
in the asymptotic Bethe ansatz.
A clean setup that allows one to study the transcendentality structure 
without admixtures from
the dressing phase is provided by single-impurity operators in the 
$\beta$-deformed $\mathcal{N}=4$ SYM 
theory.\footnote{Among other 
deformations the $\beta$-deformation is reviewed in chapter \cite{chapDeform}.}
The leading wrapping corrections have been calculated up to 
$11$ loops in \cite{Fiamberti:2008sm,Fiamberti:2008sn} and were confirmed in 
\cite{Beccaria:2009hg,Gromov:2010dy,Arutyunov:2010gu}. 
A clear pattern emerges also for the terms of lower transcendentality.
The diagrams in Figure \ref{fig:hightranswrap} are responsible for the highest 
transcendentality contribution involving $\zeta(2K-3)$.
The respective term can be traced back 
to a component $\frac{1}{2}P_K$ in the decomposition of the integrals, 
where $P_K$ is the $K$-loop cake integral given in \eqref{PK}.

\section{Conclusions}
\label{sec:concl}

We have reviewed the explicit Feynman diagram calculations 
which at small 't Hooft coupling determine the planar spectrum
of composite operators in the flavour $SU(2)$ subsector
of $\mathcal{N}=4$ SYM theory and test the 
underlying integrability.
We have presented the calculations up to two loops in detail and 
summarized the calculations and partial checks at higher loops.
The use of $\mathcal{N}=1$ superspace techniques 
and of chiral functions as operators in flavour space allowed us to  
directly interpret the Feynman diagrams in terms of the dispersion relation
and the scattering matrix that appear in the integrability-based Bethe 
ansatz.

Then, we reviewed how anomalous dimensions beyond the asymptotic limit 
can be obtained by computing the leading wrapping corrections and 
which properties and interpretation these interactions have.
The existing tests in these setups have been summarized.

\section*{Acknowledgements}

\noindent 
I am very grateful to Francesco Fiamberti and Alberto Santambrogio
for reading parts of the manuscript. 
I also want to thank Francesco Fiamberti, Matias Leoni, Andrea Mauri, 
Joseph Minahan, Alberto Santambrogio, Olof Ohlsson Sax, 
Gabriele Tartaglino-Mazzucchelli, Alessandro Torielli 
and Daniela Zanon for very pleasant collaborations on some of the papers 
reviewed here and in other chapters of this review.


\appendix

\section{\texorpdfstring{$\D$}{D}-algebra}
\label{app:Dalg}

The propagators and vertices of superfields depend not only on the bosonic, 
but also on the fermionic coordinates $\theta^\alpha$, $\bar\theta^{\dot\alpha}$,
of superspace and carry covariant 
spinor derivatives $\D_\alpha$, $\barD_{\dot\alpha}$. 
By the $\D$-algebra manipulation which consists of 
transfers, partial integrations and the use of (anti)-commutation relations 
for products of these spinor derivatives, the underlying expression is
transformed into the final result that is localized at a single point 
in the coordinates $\theta^\alpha$, $\bar\theta^{\dot\alpha}$.
We refer the reader to \cite{Gates:1983nr} for 
an introduction to the $\mathcal{N}=1$ superfield 
formalism in the adopted conventions and to \cite{Sieg:2010tz} 
for an explicit presentation of the 
relevant Feynman rules. Here, we only recall that
two $\D_\alpha$ and two $\barD_{\dot\alpha}$ have 
to remain in each loop in order to obtain a non-vanishing result. 
The loop is then localized in the fermionic coordinates. 
We indicate this by filling it grey.
Also, we recall two simple relations, $\D^2\barD^2\D^2=\Box\D^2$ and 
$\barD^2\D^2\barD^2=\Box\barD^2$, which transform spinor derivatives 
into spacetime derivatives $\Box=\partial^\mu\partial_\mu$.

The one-loop diagram \eqref{oneloopdiag} requires no $\D$-algebra
manipulations, and one directly obtains
\begin{equation}
\begin{aligned}
\settoheight{\eqoff}{$\times$}%
\setlength{\eqoff}{0.5\eqoff}%
\addtolength{\eqoff}{-27\unitlength}%
\raisebox{\eqoff}{%
\fmfframe(-2,2)(-27,2){%
\begin{fmfchar*}(50,50)
\chione
\fmfis{phantom,ptext.out=$\scriptstyle\barD^2$,ptext.clen=7,ptext.hin=2,ptext.hout=2,ptext.oin=8,ptext.oout=8,ptext.sep=;}{reverse(p1)}
\fmfis{phantom,ptext.out=$\scriptstyle\barD^2$,ptext.clen=7,ptext.hin=-10,ptext.hout=-10,ptext.oin=8,ptext.oout=8,ptext.sep=;}{reverse(p2)}
\fmfis{phantom,ptext.in=$\scriptstyle\D^2$,ptext.clen=7,ptext.hin=2,ptext.hout=2,ptext.oin=8,ptext.oout=8,ptext.sep=;}{reverse(p3)}
\fmfis{phantom,ptext.in=$\scriptstyle\barD^2$,ptext.out=$\scriptstyle\D^2$,ptext.clen=7,ptext.hin=2,ptext.hout=2,ptext.oin=8,ptext.oout=8,ptext.sep=;}{p4}
\end{fmfchar*}}}
=
\settoheight{\eqoff}{$\times$}%
\setlength{\eqoff}{0.5\eqoff}%
\addtolength{\eqoff}{-27\unitlength}%
\raisebox{\eqoff}{%
\fmfframe(-2,2)(-27,2){%
\begin{fmfchar*}(50,50)
\chione
\fmfis{phantom,ptext.out=$\scriptstyle\barD^2$,ptext.clen=7,ptext.hin=2,ptext.hout=2,ptext.oin=8,ptext.oout=8,ptext.sep=;}{reverse(p1)}
\fmfis{phantom,ptext.out=$\scriptstyle\barD^2$,ptext.clen=7,ptext.hin=-10,ptext.hout=-10,ptext.oin=8,ptext.oout=8,ptext.sep=;}{reverse(p2)}
\fmfis{phantom,ptext.in=$\scriptstyle\D^2$,ptext.clen=7,ptext.hin=2,ptext.hout=2,ptext.oin=8,ptext.oout=8,ptext.sep=;}{reverse(p3)}
\fmfcmd{fill p4--reverse(p5)--vloc(__v3)--cycle withcolor \mympostgrey ;}
\end{fmfchar*}}}
\to
-I_1
\col
\end{aligned}
\end{equation}
where the loop integral $I_1$, given in \eqref{IK} for $K=1$,
is the one extracted from the grey-scaled region. 
Its UV pole is listed in \eqref{IKpoles}.
There appears an additional factor $-1$ in front of $I_1$: 
we have to transform the full fermionic measure in the 
algebraic expression of the diagram into the chiral measure of 
the term that adds the chiral composite operator with a chiral source
to the action. This means, we replace $\de^4\theta\to\de^2\barD^2$ and 
combine the extra derivatives $\barD^2$ 
with the remaining $\D^2$ in the above diagram to $\Box$, such that  
the propagator that connects the chiral and anti-chiral cubic vertex is 
cancelled, thereby yielding the factor $-1$.
In the result we have not considered any other non-trivial prefactors of
the propagators and vertices. They are contained within the color- and flavour 
factors (chiral functions) of the complete result given in \eqref{oneloopdiag}.

The one-loop correction to the chiral vertex 
that enters \eqref{twoloopdiags} is easily evaluated
\begin{equation}
\begin{aligned}\label{ccconeloop}
\cVat[\fmfcmd{fill fullcircle scaled 10 shifted vloc(__vg1) withcolor black ;}
\fmfiv{plain,label=\small$\textcolor{white}{1}$,l.dist=0}{vloc(__vg1)}]
{plain}{plain}{plain}{phantom}{plain,ptext.in=$\scriptstyle\barD^2$}{plain,ptext.in=$\scriptstyle\barD^2$}
&=
\cVat
{plain,ptext.in=$\scriptstyle\barD^2$}{plain,ptext.in=$\scriptstyle\barD^2$,ptext.out=$\scriptstyle\D^2$}{plain,ptext.out=$\scriptstyle\D^2$}{photon}{plain,ptext.in=$\scriptstyle\barD^2$}{plain,ptext.in=$\scriptstyle\barD^2$}
+\dots
=
\left(\cVat[\fmfiset{p10}{p4--p5--p6--cycle}\fmfcmd{fill p10 withcolor \mympostgrey ;}]
{plain,l.side=left,l.dist=2,label=$\scriptstyle\Box$}{plain}{plain}{plain}{plain,ptext.in=$\scriptstyle\barD^2$}{plain,ptext.in=$\scriptstyle\barD^2$}
+\dots\right)i\lambda g_\YM\epsilon_{ijk}\tr\big(T^a\comm{T^b}{T^c}\big)
\col
\end{aligned}
\end{equation}
where the ellipsis denote the remaining two diagrams obtained by 
cyclic permutations of the external legs, and we have included the color and 
flavour factors. Also in this case, the $\Box$ 
is produced after reducing the full fermionic measure to the chiral measure
as mentioned above. 
When $\Box$ cancels the propagator a factor $-1$ is produced.

The $\D$-algebra manipulations for the diagrams 
\eqref{twoloopdiags} contributing to the
two-loop dilatation
operator are
\begin{equation}
\begin{aligned}
&
\settoheight{\eqoff}{$\times$}%
\setlength{\eqoff}{0.5\eqoff}%
\addtolength{\eqoff}{-27\unitlength}%
\raisebox{\eqoff}{%
\fmfframe(-2,2)(-27,2){%
\begin{fmfchar*}(50,50)
\chione
\fmfis{phantom,ptext.out=$\scriptstyle\barD^2$,ptext.clen=7,ptext.hin=2,ptext.hout=2,ptext.oin=8,ptext.oout=8,ptext.sep=;}{reverse(p1)}
\fmfis{phantom,ptext.out=$\scriptstyle\barD^2$,ptext.clen=7,ptext.hin=-10,ptext.hout=-10,ptext.oin=8,ptext.oout=8,ptext.sep=;}{reverse(p2)}
\fmfis{phantom,ptext.in=$\scriptstyle\D^2$,ptext.clen=7,ptext.hin=2,ptext.hout=2,ptext.oin=12,ptext.oout=8,ptext.sep=;}{reverse(p3)}
\fmfis{phantom,ptext.in=$\scriptstyle\barD^2$,ptext.out=$\scriptstyle\D^2$,ptext.clen=7,ptext.hin=2,ptext.hout=2,ptext.oin=8,ptext.oout=8,ptext.sep=;}{p4}
\fmfcmd{fill fullcircle scaled 10 shifted vloc(__vc2) withcolor black ;}
\fmfiv{plain,label=\small$\textcolor{white}{1}$,l.dist=0}{vloc(__vc2)}
\end{fmfchar*}}}
=
\settoheight{\eqoff}{$\times$}%
\setlength{\eqoff}{0.5\eqoff}%
\addtolength{\eqoff}{-27\unitlength}%
\raisebox{\eqoff}{%
\fmfframe(-2,2)(-27,2){%
\begin{fmfchar*}(50,50)
\chione
\fmfiset{p6}{subpath (0,length(p3)/2) of p3}
\fmfiset{p7}{subpath (length(p3)/2,length(p3)) of p3}
\fmfiset{p8}{subpath (0,length(p4)/2) of p4}
\fmfiset{p9}{subpath (length(p4)/2,length(p4)) of p4}
\fmfiset{p10}{subpath (0,length(p5)/2) of p5}
\fmfiset{p11}{subpath (length(p5)/2,length(p5)) of p5}
\fmfcmd{fill reverse(p11)--vm5{dir 45}..{dir 180}vm3--p7--cycle withcolor \mympostgrey ;}
\fmfi{plain}{vm5{dir 45}..{dir 180}vm3}
\fmfis{phantom,ptext.out=$\scriptstyle\barD^2$,ptext.clen=7,ptext.hin=2,ptext.hout=2,ptext.oin=8,ptext.oout=8,ptext.sep=;}{reverse(p1)}
\fmfis{phantom,ptext.out=$\scriptstyle\barD^2$,ptext.clen=7,ptext.hin=-10,ptext.hout=-10,ptext.oin=8,ptext.oout=8,ptext.sep=;}{reverse(p2)}
\fmfis{phantom,ptext.in=$\scriptstyle\D^2$,ptext.clen=7,ptext.hin=2,ptext.hout=2,ptext.oin=8,ptext.oout=8,ptext.sep=;}{reverse(p6)}
\fmfis{phantom,ptext.in=$\scriptstyle\barD^2$,ptext.out=$\scriptstyle\D^2$,ptext.clen=7,ptext.hin=2,ptext.hout=2,ptext.oin=8,ptext.oout=5,ptext.sep=;}{p4}
\fmfis{phantom,l.dist=2,label=$\scriptstyle\Box$}{p4}
\end{fmfchar*}}}
+
\settoheight{\eqoff}{$\times$}%
\setlength{\eqoff}{0.5\eqoff}%
\addtolength{\eqoff}{-27\unitlength}%
\raisebox{\eqoff}{%
\fmfframe(-2,2)(-27,2){%
\begin{fmfchar*}(50,50)
\chione
\fmfiset{p6}{subpath (0,length(p3)/2) of p3}
\fmfiset{p7}{subpath (length(p3)/2,length(p3)) of p3}
\fmfiset{p8}{subpath (0,length(p4)/2) of p4}
\fmfiset{p9}{subpath (length(p4)/2,length(p4)) of p4}
\fmfiset{p10}{subpath (0,length(p5)/2) of p5}
\fmfiset{p11}{subpath (length(p5)/2,length(p5)) of p5}
\fmfcmd{fill reverse(p9)--vm4{dir 135}..{dir 0}vm3--p7--cycle withcolor \mympostgrey ;}
\fmfi{plain}{vm4{dir 135}..{dir 0}vm3}
\fmfis{phantom,ptext.out=$\scriptstyle\barD^2$,ptext.clen=7,ptext.hin=2,ptext.hout=2,ptext.oin=8,ptext.oout=8,ptext.sep=;}{reverse(p1)}
\fmfis{phantom,ptext.out=$\scriptstyle\barD^2$,ptext.clen=7,ptext.hin=-10,ptext.hout=-10,ptext.oin=8,ptext.oout=8,ptext.sep=;}{reverse(p2)}
\fmfis{phantom,ptext.in=$\scriptstyle\D^2$,ptext.clen=7,ptext.hin=2,ptext.hout=2,ptext.oin=8,ptext.oout=8,ptext.sep=;}{reverse(p6)}
\fmfis{phantom,ptext.in=$\scriptstyle\barD^2$,ptext.out=$\scriptstyle\D^2$,ptext.clen=7,ptext.hin=2,ptext.hout=2,ptext.oin=8,ptext.oout=5,ptext.sep=;}{p8}
\fmfis{phantom,l.dist=2,label=$\scriptstyle\Box$}{p5}
\end{fmfchar*}}}
+
\settoheight{\eqoff}{$\times$}%
\setlength{\eqoff}{0.5\eqoff}%
\addtolength{\eqoff}{-27\unitlength}%
\raisebox{\eqoff}{%
\fmfframe(-2,2)(-27,2){%
\begin{fmfchar*}(50,50)
\chione
\fmfiset{p6}{subpath (0,length(p3)/2) of p3}
\fmfiset{p7}{subpath (length(p3)/2,length(p3)) of p3}
\fmfiset{p8}{subpath (0,length(p4)/2) of p4}
\fmfiset{p9}{subpath (length(p4)/2,length(p4)) of p4}
\fmfiset{p10}{subpath (0,length(p5)/2) of p5}
\fmfiset{p11}{subpath (length(p5)/2,length(p5)) of p5}
\fmfcmd{fill reverse(p9)--vm4--vm5--p11--cycle withcolor \mympostgrey ;}
\fmfi{plain}{vm4--vm5}
\fmfis{phantom,ptext.out=$\scriptstyle\barD^2$,ptext.clen=7,ptext.hin=2,ptext.hout=2,ptext.oin=8,ptext.oout=8,ptext.sep=;}{reverse(p1)}
\fmfis{phantom,ptext.out=$\scriptstyle\barD^2$,ptext.clen=7,ptext.hin=-10,ptext.hout=-10,ptext.oin=8,ptext.oout=8,ptext.sep=;}{reverse(p2)}
\fmfis{phantom,l.side=left,l.dist=2,label=$\scriptstyle\Box$,ptext.in=$\scriptstyle\D^2$,ptext.clen=7,ptext.hin=2,ptext.hout=2,ptext.oin=8,ptext.oout=8,ptext.sep=;}{reverse(p3)}
\fmfis{phantom,ptext.in=$\scriptstyle\barD^2$,ptext.out=$\scriptstyle\D^2$,ptext.clen=7,ptext.hin=2,ptext.hout=2,ptext.oin=8,ptext.oout=5,ptext.sep=;}{p8}
\end{fmfchar*}}}
=
2
\settoheight{\eqoff}{$\times$}%
\setlength{\eqoff}{0.5\eqoff}%
\addtolength{\eqoff}{-27\unitlength}%
\raisebox{\eqoff}{%
\fmfframe(-2,2)(-27,2){%
\begin{fmfchar*}(50,50)
\chione
\fmfiset{p6}{subpath (0,length(p3)/2) of p3}
\fmfiset{p7}{subpath (length(p3)/2,length(p3)) of p3}
\fmfiset{p8}{subpath (0,length(p4)/2) of p4}
\fmfiset{p9}{subpath (length(p4)/2,length(p4)) of p4}
\fmfiset{p10}{subpath (0,length(p5)/2) of p5}
\fmfiset{p11}{subpath (length(p5)/2,length(p5)) of p5}
\fmfcmd{fill reverse(p11)--vm5{dir 45}..{dir 180}vm3--p7--cycle withcolor \mympostgrey ;}
\fmfcmd{fill p4--reverse(p5)--vloc(__v3)--cycle withcolor \mympostgrey ;}
\fmfi{plain}{vm5{dir 45}..{dir 180}vm3}
\fmfis{phantom,ptext.out=$\scriptstyle\barD^2$,ptext.clen=7,ptext.hin=2,ptext.hout=2,ptext.oin=8,ptext.oout=8,ptext.sep=;}{reverse(p1)}
\fmfis{phantom,ptext.out=$\scriptstyle\barD^2$,ptext.clen=7,ptext.hin=-10,ptext.hout=-10,ptext.oin=8,ptext.oout=8,ptext.sep=;}{reverse(p2)}
\fmfis{phantom,ptext.in=$\scriptstyle\D^2$,ptext.clen=7,ptext.hin=2,ptext.hout=2,ptext.oin=8,ptext.oout=8,ptext.sep=;}{reverse(p6)}
\fmfis{phantom,l.dist=2,label=$\scriptstyle\Box$}{p4}
\end{fmfchar*}}}
\to 2I_2
\col\\
&
\settoheight{\eqoff}{$\times$}%
\setlength{\eqoff}{0.5\eqoff}%
\addtolength{\eqoff}{-27\unitlength}%
\raisebox{\eqoff}{%
\fmfframe(-2,2)(-15,2){%
\begin{fmfchar*}(50,50)
\chionetwo
\fmfis{phantom,ptext.out=$\scriptstyle\barD^2$,ptext.clen=7,ptext.hin=2,ptext.hout=2,ptext.oin=8,ptext.oout=8,ptext.sep=;}{reverse(p1)}
\fmfis{phantom,ptext.out=$\scriptstyle\barD^2$,ptext.clen=7,ptext.hin=-10,ptext.hout=-10,ptext.oin=8,ptext.oout=8,ptext.sep=;}{reverse(p2)}
\fmfis{phantom,ptext.in=$\scriptstyle\D^2$,ptext.clen=7,ptext.hin=2,ptext.hout=2,ptext.oin=8,ptext.oout=8,ptext.sep=;}{reverse(p3)}
\fmfis{phantom,ptext.in=$\scriptstyle\barD^2$,ptext.out=$\scriptstyle\D^2$,ptext.clen=7,ptext.hin=2,ptext.hout=2,ptext.oin=8,ptext.oout=8,ptext.sep=;}{p4}
\fmfis{phantom,ptext.out=$\scriptstyle\barD^2$,ptext.clen=7,ptext.hin=2,ptext.hout=2,ptext.oin=8,ptext.oout=8,ptext.sep=;}{reverse(p5)}
\fmfis{phantom,ptext.out=$\scriptstyle\barD^2$,ptext.clen=7,ptext.hin=-10,ptext.hout=-10,ptext.oin=8,ptext.oout=8,ptext.sep=;}{reverse(p6)}
\fmfis{phantom,ptext.in=$\scriptstyle\D^2$,ptext.clen=7,ptext.hin=2,ptext.hout=2,ptext.oin=8,ptext.oout=8,ptext.sep=;}{reverse(p7)}
\fmfis{phantom,ptext.in=$\scriptstyle\barD^2$,ptext.out=$\scriptstyle\D^2$,ptext.clen=7,ptext.hin=2,ptext.hout=2,ptext.oin=8,ptext.oout=8,ptext.sep=;}{p8}
\end{fmfchar*}}}
=
\settoheight{\eqoff}{$\times$}%
\setlength{\eqoff}{0.5\eqoff}%
\addtolength{\eqoff}{-27\unitlength}%
\raisebox{\eqoff}{%
\fmfframe(-2,2)(-15,2){%
\begin{fmfchar*}(50,50)
\chionetwo
\fmfis{phantom,ptext.out=$\scriptstyle\barD^2$,ptext.clen=7,ptext.hin=2,ptext.hout=2,ptext.oin=8,ptext.oout=8,ptext.sep=;}{reverse(p1)}
\fmfis{phantom,ptext.out=$\scriptstyle\barD^2$,ptext.clen=7,ptext.hin=-10,ptext.hout=-10,ptext.oin=8,ptext.oout=8,ptext.sep=;}{reverse(p2)}
\fmfis{phantom,ptext.in=$\scriptstyle\D^2$,ptext.clen=7,ptext.hin=2,ptext.hout=2,ptext.oin=8,ptext.oout=8,ptext.sep=;}{reverse(p3)}
\fmfis{phantom,l.side=left,l.dist=2,label=$\scriptstyle\Box$,ptext.clen=7,ptext.hin=2,ptext.hout=2,ptext.oin=8,ptext.oout=8,ptext.sep=;}{reverse(p5)}
\fmfis{phantom,ptext.out=$\scriptstyle\barD^2$,ptext.clen=7,ptext.hin=-10,ptext.hout=-10,ptext.oin=8,ptext.oout=8,ptext.sep=;}{reverse(p6)}
\fmfcmd{fill reverse(p4)--reverse(p5)--p7--p8--vloc(__v5)--vloc(__v4)--cycle withcolor \mympostgrey ;}
\fmfcmd{fill reverse(p8)--vloc(__v5)--vloc(__v6)--reverse(p9)--cycle withcolor \mympostgrey ;}
\end{fmfchar*}}}
\to I_2
\col
\end{aligned}
\end{equation}
where equalities hold up to disregarded finite contributions, 
and the final expressions 
in terms of the integral $I_2$ consider the aforementioned factor $-1$.

\unitlength=0.875mm

\section{Integrals}
\label{app:integrals}

Using the scalar $G$-function defined as
\begin{equation}
G(\alpha,\beta)=
\frac{\Gamma(\tfrac{D}{2}-\alpha)\Gamma(\tfrac{D}{2}-\beta)\Gamma(\alpha+\beta-\tfrac{D}{2})}{(4\pi)^{\frac{D}{2}}\Gamma(\alpha)\Gamma(\beta)\Gamma(D-\alpha-\beta)}\col
\end{equation}
in $D$-dimensional Euclidean space,
the following integrals can be found exactly to all loop orders
\begin{equation}\label{IK}
\begin{aligned}
I_K=
\settoheight{\eqoff}{$\times$}%
\setlength{\eqoff}{0.5\eqoff}%
\addtolength{\eqoff}{-13\unitlength}%
\raisebox{\eqoff}{%
\fmfframe(-6,-2)(-1,-2){%
\begin{fmfchar*}(30,30)
  \fmfleft{in}
  \fmfright{out1}
\fmf{phantom}{in,v1}
\fmf{phantom}{out,v2}
\fmfforce{(0,0.5h)}{in}
\fmfforce{(w,0.5h)}{out}
\fmfpoly{phantom}{v1,va4,va3,v2,va2,va1}
\fmffixed{(0.75w,0)}{v1,v2}
\fmf{phantom}{vc,v1}
\fmf{plain}{vc,v2}
\fmffreeze
\fmf{phantom}{v1,va1}
\fmf{plain}{va1,va2}
\fmf{plain}{va2,v2}
\fmf{dashes}{v2,va3}
\fmf{plain}{va3,va4}
\fmf{plain}{vc,va1}
\fmf{plain}{vc,va2}
\fmf{dashes}{vc,va3}
\fmf{plain}{vc,va4}
\fmffreeze
\fmf{plain,left=0.5}{vc,va1}
\fmfv{l=$\scriptscriptstyle 1$,l.dist=2}{va1}
\fmfv{l=$\scriptscriptstyle 2$,l.dist=2}{va2}
\fmfv{l=$\scriptscriptstyle 3$,l.dist=2}{v2}
\fmfv{l=$\scriptscriptstyle K-1$,l.dist=2}{va3}
\fmfv{l=$\scriptscriptstyle K$,l.dist=2}{va4}
\end{fmfchar*}}}
=\prod_{k=0}^{K-1}G(1-(\tfrac{D}{2}-2)k,1)
\pnt
\end{aligned}
\end{equation}
They are logarithmically divergent in $D=4-2\varepsilon$ dimensions, and 
their overall UV divergence is obtained with the operations $\Kop$ to extract the pole part and $\Rop$ to subtract subdivergences as
\begin{equation}
\Kop\Rop(I_K)=\Kop\Big(I_K-\sum_{k=1}^{K-1}\Kop\Rop(I_k)I_{K-k}\Big)
\pnt
\end{equation}
To the first few loop orders, one finds
\begin{equation}\label{IKpoles}
\begin{aligned}
\Kop\Rop(I_1)&=\frac{1}{(4\pi)^2}\frac{1}{\varepsilon} \col\\
\Kop\Rop(I_2)&=\frac{1}{(4\pi)^4}\Big(-\frac{1}{2\varepsilon^2}+\frac{1}{2\varepsilon}\Big) \col\\
\Kop\Rop(I_3)&=\frac{1}{(4\pi)^6}\Big(\frac{1}{6\varepsilon^3}-\frac{1}{2\varepsilon^2}+\frac{2}{3\varepsilon}\Big) \col\\
\Kop\Rop(I_4)&=\frac{1}{(4\pi)^8}\Big(-\frac{1}{24\varepsilon^4}+\frac{1}{4\varepsilon^3}-\frac{19}{24\varepsilon^2}+\frac{5}{4\varepsilon}\Big) \col\\
\Kop\Rop(I_5)&=\frac{1}{(4\pi)^{10}}\Big(\frac{1}{120\varepsilon^5}-\frac{1}{12\varepsilon^4}+\frac{11}{24\varepsilon^3}-\frac{19}{12\varepsilon^2}+\frac{14}{5\varepsilon}\Big) \col\\
\Kop\Rop(I_6)&=\frac{1}{(4\pi)^{12}}\Big(-\frac{1}{720\varepsilon^6}+\frac{1}{48\varepsilon^5}-\frac{25}{144\varepsilon^4}+\frac{47}{48\varepsilon^3}-\frac{1313}{360\varepsilon^2}+\frac{7}{\varepsilon}\Big) \pnt\\
\end{aligned}
\end{equation}

The pole parts of the integrals 
that appear in the calculations of the four-loop dressing
phase or of the wrapping interactions at critical wrapping order 
can very efficiently be computed by using a modified and extended 
version of the Gegenbauer polynomial $x$-space technique 
\cite{Chetyrkin:1980pr,Kotikov:1995cw,Fiamberti:2008sh}.
The integral of the
simplest contribution that allows us to determine the leading 
four-loop coefficient of the 
dressing phase reads
\begin{equation}
\begin{aligned}\label{Ipoles}
I_\beta&=
\settoheight{\eqoff}{$\times$}%
\setlength{\eqoff}{0.5\eqoff}%
\addtolength{\eqoff}{-7.5\unitlength}%
\raisebox{\eqoff}{%
\begin{fmfchar*}(20,15)
\fmfleft{in}
\fmfright{out}
\fmf{plain}{in,v1}
\fmf{plain,left=0.25}{v1,v2}
\fmf{plain,left=0.25}{v2,v3}
\fmf{plain,left=0.25}{v3,v4}
\fmf{plain,left=0.25}{v4,v1}
\fmf{plain,tension=0.5,right=0.25}{v1,v0,v1}
\fmf{phantom}{v0,v3}
\fmf{plain}{v2,v0}
\fmf{plain}{v0,v4}
\fmf{plain}{v3,out}
\fmffixed{(0.9w,0)}{v1,v3}
\fmffixed{(0,0.45w)}{v4,v2}
\fmffreeze
\end{fmfchar*}}
\col\quad
\Kop\Rop(I_\beta)
=\frac{1}{(4\pi)^8}\Big(
-\frac{1}{12\varepsilon^4}+\frac{1}{3\varepsilon^3}
-\frac{5}{12\varepsilon^2}
-\frac{1}{\varepsilon}\Big(\frac{1}{2}-\zeta(3)\Big)\Big)
\pnt
\end{aligned}
\end{equation}

The terms of highest transcendentality from wrapping corrections at critical 
order are determined by the cake integral. This integral is logarithmically 
divergent for $K\ge3$ loops and reads
\begin{equation}\label{PK}
P_K=
\settoheight{\eqoff}{$\times$}%
\setlength{\eqoff}{0.5\eqoff}%
\addtolength{\eqoff}{-13\unitlength}%
\raisebox{\eqoff}{%
\fmfframe(-1,-2)(-1,-2){%
\begin{fmfchar*}(30,30)
  \fmfleft{in}
  \fmfright{out1}
\fmf{phantom}{in,v1}
\fmf{phantom}{out,v2}
\fmfforce{(0,0.5h)}{in}
\fmfforce{(w,0.5h)}{out}
\fmfpoly{phantom}{v1,va4,va3,v2,va2,va1}
\fmffixed{(0.75w,0)}{v1,v2}
\fmf{phantom}{vc,v1}
\fmf{plain}{vc,v2}
\fmffreeze
\fmf{plain}{v1,va1}
\fmf{plain}{va1,va2}
\fmf{plain}{va2,v2}
\fmf{plain}{v2,va3}
\fmf{dashes}{va3,va4}
\fmf{plain}{va4,v1}
\fmf{plain}{vc,va1}
\fmf{plain}{vc,va2}
\fmf{plain}{vc,va3}
\fmf{dashes}{vc,va4}
\fmf{plain}{vc,v1}
\fmffreeze
\fmfv{l=$\scriptscriptstyle 1$,l.dist=2}{va1}
\fmfv{l=$\scriptscriptstyle 2$,l.dist=2}{va2}
\fmfv{l=$\scriptscriptstyle 3$,l.dist=2}{v2}
\fmfv{l=$\scriptscriptstyle 4$,l.dist=2}{va3}
\fmfv{l=$\scriptscriptstyle K-1$,l.dist=2}{va4}
\fmfv{l=$\scriptscriptstyle K$,l.dist=2}{v1}
\end{fmfchar*}}}
\col\qquad
\Kop(P_K)
=\frac{1}{(4\pi)^{2K}}\frac{1}{\varepsilon}
\frac{2}{K}\binom{2K-3}{K-1}\zeta(2K-3)
\col
\end{equation}
where the pole part has been obtained in \cite{Broadhurst:1985vq}
at generic loop order.

\phantomsection
\addcontentsline{toc}{section}{\refname}

\begin{thebibliography}{IV.3}
\ifx\href\asklfhas\newcommand{\href}[2]{#2}\fi
\ifx\arxivref\asklfhas\newcommand{\arxivref}[2]{\href{http://arxiv.org/abs/#1}%
{#2}}\fi
\ifx\doiref\asklfhas\newcommand{\doiref}[2]{\href{http://dx.doi.org/#1}{#2}}\fi
\raggedright
\small
\parskip 0pt

\bibitem[I.1]{chapChain}
J.~Minahan,
\textit{``Review of AdS/CFT Integrability, Chapter I.1: Spin Chains in
  $\mathcal{N}$ = 4 SYM''},
\texttt{\arxivref{1012.3983}{arxiv:1012.3983}}.

\bibitem[I.3]{chapLR}
A.~Rej,
\textit{``Review of AdS/CFT Integrability, Chapter I.3: Long-range spin
  chains''},
\texttt{\arxivref{1012.3985}{arxiv:1012.3985}}.

\bibitem[II.1]{chapSpinning}
A.~Tseytlin,
\textit{``Review of AdS/CFT Integrability, Chapter II.1: Classical $AdS_5\times
  S^5$ string solutions''},
\texttt{\arxivref{1012.3986}{arxiv:1012.3986}}.

\bibitem[II.2]{chapQstring}
T.~McLoughlin,
\textit{``Review of AdS/CFT Integrability, Chapter II.2: Quantum Strings in
  $AdS_5\times S^5$''},
\texttt{\arxivref{1012.3987}{arxiv:1012.3987}}.

\bibitem[III.1]{chapABA}
M.~Staudacher,
\textit{``Review of AdS/CFT Integrability, Chapter III.1: Bethe Ans\"atze and
  the R-Matrix Formalism''},
\texttt{\arxivref{1012.3990}{arxiv:1012.3990}}.

\bibitem[III.3]{chapSProp}
P.~Vieira and D.~Volin,
\textit{``Review of AdS/CFT Integrability, Chapter III.3: The dressing
  factor''},
\texttt{\arxivref{1012.3992}{arxiv:1012.3992}}.

\bibitem[III.4]{chapTwist}
L.~Freyhult,
\textit{``Review of AdS/CFT Integrability, Chapter III.4: Twist states and the
  cusp anomalous dimension''},
\texttt{\arxivref{1012.3993}{arxiv:1012.3993}}.

\bibitem[III.5]{chapLuescher}
R.~Janik,
\textit{``Review of AdS/CFT Integrability, Chapter III.5: L\"uscher
  corrections''},
\texttt{\arxivref{1012.3994}{arxiv:1012.3994}}.

\bibitem[III.6]{chapTBA}
Z.~Bajnok,
\textit{``Review of AdS/CFT Integrability, Chapter III.6: Thermodynamic Bethe
  Ansatz''},
\texttt{\arxivref{1012.3995}{arxiv:1012.3995}}.

\bibitem[III.7]{chapTrans}
V.~Kazakov and N.~Gromov,
\textit{``Review of AdS/CFT Integrability, Chapter III.7: Hirota Dynamics for
  Quantum Integrability''},
\texttt{\arxivref{1012.3996}{arxiv:1012.3996}}.

\bibitem[IV.1]{chapObserv}
C.~Kristjansen,
\textit{``Review of AdS/CFT Integrability, Chapter IV.1: Aspects of
  Non-planarity''},
\texttt{\arxivref{1012.3997}{arxiv:1012.3997}}.

\bibitem[IV.2]{chapDeform}
K.~Zoubos,
\textit{``Review of AdS/CFT Integrability, Chapter IV.2: Deformations,
  Orbifolds and Open Boundaries''},
\texttt{\arxivref{1012.3998}{arxiv:1012.3998}}.

\bibitem[IV.3]{chapN6}
T.~Klose,
\textit{``Review of AdS/CFT Integrability, Chapter IV.3: $\mathcal{N}$ = 6
  Chern-Simons and Strings on $AdS_4 \times CP^3$''},
\texttt{\arxivref{1012.3999}{arxiv:1012.3999}}.

\bibitem{Maldacena:1997re}
J.~M.~Maldacena,
\textit{``{The large $N$ limit of superconformal field theories and
  supergravity}''},
\textsf{Adv.~Theor.~Math.~Phys.~2,~231~(1998)},
\texttt{\arxivref{hep-th/9711200}{hep-th/9711200}}.

\bibitem{Gubser:1998bc}
S.~S.~Gubser, I.~R.~Klebanov and A.~M.~Polyakov,
\textit{``{Gauge theory correlators from non-critical string theory}''},
\textsf{\doiref{10.1016/S0370-2693(98)00377-3}{Phys.~Lett.~B428,~105~(1998)}},
\texttt{\arxivref{hep-th/9802109}{hep-th/9802109}}.

\bibitem{Witten:1998qj}
E.~Witten,
\textit{``{Anti-de Sitter space and holography}''},
\textsf{Adv.~Theor.~Math.~Phys.~2,~253~(1998)},
\texttt{\arxivref{hep-th/9802150}{hep-th/9802150}}.

\bibitem{Serban:2004jf}
D.~Serban and M.~Staudacher,
\textit{``{Planar $\mathcal{N}=4$ gauge theory and the Inozemtsev long range
  spin chain}''},
\textsf{\doiref{10.1088/1126-6708/2004/06/001}{JHEP~0406,~001~(2004)}},
\texttt{\arxivref{hep-th/0401057}{hep-th/0401057}}.

\bibitem{Beisert:2004hm}
N.~Beisert, V.~Dippel and M.~Staudacher,
\textit{``{A novel long range spin chain and planar $\mathcal{N}=4$ super
  Yang-Mills}''},
\textsf{\doiref{10.1088/1126-6708/2004/07/075}{JHEP~0407,~075~(2004)}},
\texttt{\arxivref{hep-th/0405001}{hep-th/0405001}}.

\bibitem{Sieg:2005kd}
C.~Sieg and A.~Torrielli,
\textit{``{Wrapping interactions and the genus expansion of the $2$- point
  function of composite operators}''},
\textsf{\doiref{10.1016/j.nuclphysb.2005.06.011}{Nucl.~Phys.~B723,~3~(2005)}},
\texttt{\arxivref{hep-th/0505071}{hep-th/0505071}}.

\bibitem{Ambjorn:2005wa}
J.~Ambjorn, R.~A.~Janik and C.~Kristjansen,
\textit{``{Wrapping interactions and a new source of corrections to the
  spin-chain / string duality}''},
\textsf{\doiref{10.1016/j.nuclphysb.2005.12.007}{Nucl.~Phys.~B736,~288~(2006)}%
},
\texttt{\arxivref{hep-th/0510171}{hep-th/0510171}}.

\bibitem{Siegel:1979wq}
W.~Siegel,
\textit{``{Supersymmetric Dimensional Regularization via Dimensional
  Reduction}''},
\textsf{\doiref{10.1016/0370-2693(79)90282-X}{Phys.~Lett.~B84,~193~(1979)}}.

\bibitem{Grisaru:1980jc}
M.~T.~Grisaru, M.~Rocek and W.~Siegel,
\textit{``{Superloops 3, beta 0: A Calculation in $\mathcal{N}=4$ Yang-Mills
  Theory}''},
\textsf{\doiref{10.1016/0550-3213(81)90550-2}{Nucl.~Phys.~B183,~141~(1981)}}.

\bibitem{Grisaru:1980nk}
M.~T.~Grisaru, M.~Rocek and W.~Siegel,
\textit{``{Zero Three Loop beta Function in $\mathcal{N}=4$ Superyang-Mills
  Theory}''},
\textsf{\doiref{10.1103/PhysRevLett.45.1063}{Phys.~Rev.~Lett.~45,~1063~(1980)}%
}.

\bibitem{Caswell:1980yi}
W.~E.~Caswell and D.~Zanon,
\textit{``{Vanishing Three Loop beta Function in $\mathcal{N}=4$ Supersymmetric
  Yang-Mills Theory}''},
\textsf{\doiref{10.1016/0370-2693(81)90764-4}{Phys.~Lett.~B100,~152~(1981)}}.

\bibitem{Caswell:1980ru}
W.~E.~Caswell and D.~Zanon,
\textit{``{Zero Three Loop beta Function in the $\mathcal{N}=4$ Supersymmetric
  Yang-Mills Theory}''},
\textsf{\doiref{10.1016/0550-3213(81)90461-2}{Nucl.~Phys.~B182,~125~(1981)}}.

\bibitem{Mandelstam:1982cb}
S.~Mandelstam,
\textit{``{Light Cone Superspace and the Ultraviolet Finiteness of the
  $\mathcal{N}=4$ Model}''},
\textsf{\doiref{10.1016/0550-3213(83)90179-7}{Nucl.~Phys.~B213,~149~(1983)}}.

\bibitem{Brink:1982wv}
L.~Brink, O.~Lindgren and B.~E.~W.~Nilsson,
\textit{``{The Ultraviolet Finiteness of the $\mathcal{N}=4$ Yang-Mills
  Theory}''},
\textsf{\doiref{10.1016/0370-2693(83)91210-8}{Phys.~Lett.~B123,~323~(1983)}}.

\bibitem{Howe:1982tm}
P.~S.~Howe, K.~S.~Stelle and P.~K.~Townsend,
\textit{``{The Relaxed Hypermultiplet: An Unconstrained $\mathcal{N}=2$
  Superfield Theory}''},
\textsf{\doiref{10.1016/0550-3213(83)90249-3}{Nucl.~Phys.~B214,~519~(1983)}}.

\bibitem{Howe:1983sr}
P.~S.~Howe, K.~S.~Stelle and P.~K.~Townsend,
\textit{``{Miraculous Ultraviolet Cancellations in Supersymmetry Made
  Manifest}''},
\textsf{\doiref{10.1016/0550-3213(84)90528-5}{Nucl.~Phys.~B236,~125~(1984)}}.

\bibitem{Ferrara:1974pu}
S.~Ferrara and B.~Zumino,
\textit{``{Supergauge Invariant Yang-Mills Theories}''},
\textsf{\doiref{10.1016/0550-3213(74)90559-8}{Nucl.~Phys.~B79,~413~(1974)}}.

\bibitem{Sieg:2010tz}
C.~Sieg,
\textit{``{Superspace calculation of the three-loop dilatation operator of
  $\mathcal{N}=4$ SYM theory}''},
\texttt{\arxivref{1008.3351}{arxiv:1008.3351}}.

\bibitem{Gates:1983nr}
S.~J.~Gates, M.~T.~Grisaru, M.~Rocek and W.~Siegel,
\textit{``{Superspace, or one thousand and one lessons in supersymmetry}''},
\textsf{Front.~Phys.~58,~1~(1983)},
\texttt{\arxivref{hep-th/0108200}{hep-th/0108200}}.

\bibitem{Minahan:2005jq}
J.~A.~Minahan,
\textit{``{The $SU(2)$ sector in AdS/CFT}''},
\textsf{\doiref{10.1002/prop.200410204}{Fortsch.~Phys.~53,~828~(2005)}},
\texttt{\arxivref{hep-th/0503143}{hep-th/0503143}}.

\bibitem{Beisert:2003tq}
N.~Beisert, C.~Kristjansen and M.~Staudacher,
\textit{``{The dilatation operator of $\mathcal{N}=4$ super Yang-Mills
  theory}''},
\textsf{\doiref{10.1016/S0550-3213(03)00406-1}{Nucl.~Phys.~B664,~131~(2003)}},
\texttt{\arxivref{hep-th/0303060}{hep-th/0303060}}.

\bibitem{Beisert:2005wv}
N.~Beisert and T.~Klose,
\textit{``{Long-range $gl(n)$ integrable spin chains and plane-wave matrix
  theory}''},
\textsf{J.~Stat.~Mech.~0607,~P006~(2006)},
\texttt{\arxivref{hep-th/0510124}{hep-th/0510124}}.

\bibitem{Fiamberti:2007rj}
F.~Fiamberti, A.~Santambrogio, C.~Sieg and D.~Zanon,
\textit{``{Wrapping at four loops in $\mathcal{N}=4$ SYM}''},
\textsf{\doiref{10.1016/j.physletb.2008.06.061}{Phys.~Lett.~B666,~100~(2008)}},
\texttt{\arxivref{0712.3522}{arxiv:0712.3522}}.

\bibitem{Fiamberti:2008sh}
F.~Fiamberti, A.~Santambrogio, C.~Sieg and D.~Zanon,
\textit{``{Anomalous dimension with wrapping at four loops in $\mathcal{N}=4$
  SYM}''},
\textsf{\doiref{10.1016/j.nuclphysb.2008.07.014}{Nucl.~Phys.~B805,~231~(2008)}%
},
\texttt{\arxivref{0806.2095}{arxiv:0806.2095}}.

\bibitem{Berenstein:2002jq}
D.~E.~Berenstein, J.~M.~Maldacena and H.~S.~Nastase,
\textit{``{Strings in flat space and pp waves from $\mathcal{N}=4$ super Yang
  Mills}''},
\textsf{\doiref{10.1088/1126-6708/2002/04/013}{JHEP~0204,~013~(2002)}},
\texttt{\arxivref{hep-th/0202021}{hep-th/0202021}}.

\bibitem{Minahan:2002ve}
J.~A.~Minahan and K.~Zarembo,
\textit{``{The Bethe-ansatz for $\mathcal{N}=4$ super Yang-Mills}''},
\textsf{\doiref{10.1088/1126-6708/2003/03/013}{JHEP~0303,~013~(2003)}},
\texttt{\arxivref{hep-th/0212208}{hep-th/0212208}}.

\bibitem{Gross:2002mh}
D.~J.~Gross, A.~Mikhailov and R.~Roiban,
\textit{``{A calculation of the plane wave string Hamiltonian from
  $\mathcal{N}=4$ super-Yang-Mills theory}''},
\textsf{\doiref{10.1088/1126-6708/2003/05/025}{JHEP~0305,~025~(2003)}},
\texttt{\arxivref{hep-th/0208231}{hep-th/0208231}}.

\bibitem{Eden:2004ua}
B.~Eden, C.~Jarczak and E.~Sokatchev,
\textit{``{A three-loop test of the dilatation operator in $\mathcal{N}=4$
  SYM}''},
\textsf{\doiref{10.1016/j.nuclphysb.2005.01.036}{Nucl.~Phys.~B712,~157~(2005)}%
},
\texttt{\arxivref{hep-th/0409009}{hep-th/0409009}}.

\bibitem{Anselmi:1998ms}
D.~Anselmi,
\textit{``{The $\mathcal{N}=4$ quantum conformal algebra}''},
\textsf{\doiref{10.1016/S0550-3213(98)00848-7}{Nucl.~Phys.~B541,~369~(1999)}},
\texttt{\arxivref{hep-th/9809192}{hep-th/9809192}}.

\bibitem{Beisert:2005tm}
N.~Beisert,
\textit{``{The $su(2|2)$ dynamic S-matrix}''},
\textsf{Adv.~Theor.~Math.~Phys.~12,~945~(2008)},
\texttt{\arxivref{hep-th/0511082}{hep-th/0511082}}.

\bibitem{Gross:2002su}
D.~J.~Gross, A.~Mikhailov and R.~Roiban,
\textit{``{Operators with large R charge in $\mathcal{N}=4$ Yang-Mills
  theory}''},
\textsf{\doiref{10.1006/aphy.2002.6293}{Annals~Phys.~301,~31~(2002)}},
\texttt{\arxivref{hep-th/0205066}{hep-th/0205066}}.

\bibitem{Beisert:2006ez}
N.~Beisert, B.~Eden and M.~Staudacher,
\textit{``{Transcendentality and crossing}''},
\textsf{J.~Stat.~Mech.~0701,~P021~(2007)},
\texttt{\arxivref{hep-th/0610251}{hep-th/0610251}}.

\bibitem{Arutyunov:2004vx}
G.~Arutyunov, S.~Frolov and M.~Staudacher,
\textit{``{Bethe ansatz for quantum strings}''},
\textsf{\doiref{10.1088/1126-6708/2004/10/016}{JHEP~0410,~016~(2004)}},
\texttt{\arxivref{hep-th/0406256}{hep-th/0406256}}.

\bibitem{Hernandez:2006tk}
R.~Hernandez and E.~Lopez,
\textit{``{Quantum corrections to the string Bethe ansatz}''},
\textsf{\doiref{10.1088/1126-6708/2006/07/004}{JHEP~0607,~004~(2006)}},
\texttt{\arxivref{hep-th/0603204}{hep-th/0603204}}.

\bibitem{Bern:2006ew}
Z.~Bern, M.~Czakon, L.~J.~Dixon, D.~A.~Kosower and V.~A.~Smirnov,
\textit{``{The Four-Loop Planar Amplitude and Cusp Anomalous Dimension in
  Maximally Supersymmetric Yang-Mills Theory}''},
\textsf{\doiref{10.1103/PhysRevD.75.085010}{Phys.~Rev.~D75,~085010~(2007)}},
\texttt{\arxivref{hep-th/0610248}{hep-th/0610248}}.

\bibitem{Santambrogio:2002sb}
A.~Santambrogio and D.~Zanon,
\textit{``{Exact anomalous dimensions of $\mathcal{N}=4$ Yang-Mills operators
  with large R charge}''},
\textsf{\doiref{10.1016/S0370-2693(02)02627-8}{Phys.~Lett.~B545,~425~(2002)}},
\texttt{\arxivref{hep-th/0206079}{hep-th/0206079}}.

\bibitem{Eden:2005bt}
B.~Eden,
\textit{``{A two-loop test for the factorised S-matrix of planar
  $\mathcal{N}=4$}''},
\textsf{\doiref{10.1016/j.nuclphysb.2006.01.012}{Nucl.~Phys.~B738,~409~(2006)}%
},
\texttt{\arxivref{hep-th/0501234}{hep-th/0501234}}.

\bibitem{Beisert:2007hz}
N.~Beisert, T.~McLoughlin and R.~Roiban,
\textit{``{The Four-Loop Dressing Phase of $\mathcal{N}=4$ SYM}''},
\textsf{\doiref{10.1103/PhysRevD.76.046002}{Phys.~Rev.~D76,~046002~(2007)}},
\texttt{\arxivref{0705.0321}{arxiv:0705.0321}}.

\bibitem{Fiamberti:2009jw}
F.~Fiamberti, A.~Santambrogio and C.~Sieg,
\textit{``{Five-loop anomalous dimension at critical wrapping order in
  $\mathcal{N}=4$ SYM}''},
\texttt{\arxivref{0908.0234}{arxiv:0908.0234}}.

\bibitem{Konishi:1983hf}
K.~Konishi,
\textit{``{Anomalous Supersymmetry Transformation of Some Composite Operators
  in SQCD}''},
\textsf{\doiref{10.1016/0370-2693(84)90311-3}{Phys.~Lett.~B135,~439~(1984)}}.

\bibitem{Anselmi:1996mq}
D.~Anselmi, M.~T.~Grisaru and A.~Johansen,
\textit{``{A Critical Behaviour of Anomalous Currents, Electric- Magnetic
  Universality and CFT$_4$}''},
\textsf{\doiref{10.1016/S0550-3213(97)00108-9}{Nucl.~Phys.~B491,~221~(1997)}},
\texttt{\arxivref{hep-th/9601023}{hep-th/9601023}}.

\bibitem{Anselmi:1996dd}
D.~Anselmi, D.~Z.~Freedman, M.~T.~Grisaru and A.~A.~Johansen,
\textit{``{Universality of the operator product expansions of SCFT(4)}''},
\textsf{\doiref{10.1016/S0370-2693(97)00007-5}{Phys.~Lett.~B394,~329~(1997)}},
\texttt{\arxivref{hep-th/9608125}{hep-th/9608125}}.

\bibitem{Bianchi:1999ge}
M.~Bianchi, S.~Kovacs, G.~Rossi and Y.~S.~Stanev,
\textit{``{On the logarithmic behavior in $\mathcal{N}=4$ SYM theory}''},
\textsf{\doiref{10.1088/1126-6708/1999/08/020}{JHEP~9908,~020~(1999)}},
\texttt{\arxivref{hep-th/9906188}{hep-th/9906188}}.

\bibitem{Bianchi:2000hn}
M.~Bianchi, S.~Kovacs, G.~Rossi and Y.~S.~Stanev,
\textit{``{Anomalous dimensions in $\mathcal{N}=4$ SYM theory at order
  g**4}''},
\textsf{\doiref{10.1016/S0550-3213(00)00312-6}{Nucl.~Phys.~B584,~216~(2000)}},
\texttt{\arxivref{hep-th/0003203}{hep-th/0003203}}.

\bibitem{Arutyunov:2001mh}
G.~Arutyunov, B.~Eden, A.~C.~Petkou and E.~Sokatchev,
\textit{``{Exceptional non-renormalization properties and OPE analysis of
  chiral four-point functions in $\mathcal{N}=4$ SYM(4)}''},
\textsf{\doiref{10.1016/S0550-3213(01)00569-7}{Nucl.~Phys.~B620,~380~(2002)}},
\texttt{\arxivref{hep-th/0103230}{hep-th/0103230}}.

\bibitem{Bianchi:2001cm}
M.~Bianchi, S.~Kovacs, G.~Rossi and Y.~S.~Stanev,
\textit{``{Properties of the Konishi multiplet in $\mathcal{N}=4$ SYM
  theory}''},
\textsf{\doiref{10.1088/1126-6708/2001/05/042}{JHEP~0105,~042~(2001)}},
\texttt{\arxivref{hep-th/0104016}{hep-th/0104016}}.

\bibitem{Kotikov:2003fb}
A.~V.~Kotikov, L.~N.~Lipatov and V.~N.~Velizhanin,
\textit{``{Anomalous dimensions of Wilson operators in $\mathcal{N}=4$ SYM
  theory}''},
\textsf{\doiref{10.1016/S0370-2693(03)00184-9}{Phys.~Lett.~B557,~114~(2003)}},
\texttt{\arxivref{hep-ph/0301021}{hep-ph/0301021}}.

\bibitem{Kotikov:2004er}
A.~V.~Kotikov, L.~N.~Lipatov, A.~I.~Onishchenko and V.~N.~Velizhanin,
\textit{``{Three-loop universal anomalous dimension of the Wilson operators in
  $\mathcal{N}=4$ SUSY Yang-Mills model}''},
\textsf{\doiref{10.1016/j.physletb.2004.05.078}{Phys.~Lett.~B595,~521~(2004)}},
\texttt{\arxivref{hep-th/0404092}{hep-th/0404092}}.

\bibitem{Moch:2004pa}
S.~Moch, J.~A.~M.~Vermaseren and A.~Vogt,
\textit{``{The three-loop splitting functions in QCD: The non-singlet case}''},
\textsf{\doiref{10.1016/j.nuclphysb.2004.03.030}{Nucl.~Phys.~B688,~101~(2004)}%
},
\texttt{\arxivref{hep-ph/0403192}{hep-ph/0403192}}.

\bibitem{Kotikov:2002ab}
A.~V.~Kotikov and L.~N.~Lipatov,
\textit{``{DGLAP and BFKL evolution equations in the $\mathcal{N}=4$
  supersymmetric gauge theory}''},
\textsf{Nucl.~Phys.~B661,~19~(2003)},
\texttt{\arxivref{hep-ph/0208220}{hep-ph/0208220}}.

\bibitem{Fiamberti:2010fw}
F.~Fiamberti, A.~Santambrogio and C.~Sieg,
\textit{``{Superspace methods for the computation of wrapping effects in the
  standard and beta-deformed $\mathcal{N}=4$ SYM}''},
\texttt{\arxivref{1006.3475}{arxiv:1006.3475}}.

\bibitem{Fischbacher:2004iu}
T.~Fischbacher, T.~Klose and J.~Plefka,
\textit{``{Planar plane-wave matrix theory at the four loop order:
  Integrability without BMN scaling}''},
\textsf{\doiref{10.1088/1126-6708/2005/02/039}{JHEP~0502,~039~(2005)}},
\texttt{\arxivref{hep-th/0412331}{hep-th/0412331}}.

\bibitem{Bajnok:2008bm}
Z.~Bajnok and R.~A.~Janik,
\textit{``{Four-loop perturbative Konishi from strings and finite size effects
  for multiparticle states}''},
\textsf{\doiref{10.1016/j.nuclphysb.2008.08.020}{Nucl.~Phys.~B807,~625~(2009)}%
},
\texttt{\arxivref{0807.0399}{arxiv:0807.0399}}.

\bibitem{Velizhanin:2008jd}
V.~N.~Velizhanin,
\textit{``{The Four-Loop Konishi in $\mathcal{N}=4$ SYM}''},
\texttt{\arxivref{0808.3832}{arxiv:0808.3832}}.

\bibitem{Gromov:2009bc}
N.~Gromov, V.~Kazakov, A.~Kozak and P.~Vieira,
\textit{``{Integrability for the Full Spectrum of Planar AdS/CFT II}''},
\texttt{\arxivref{0902.4458}{arxiv:0902.4458}}.

\bibitem{Gromov:2009tv}
N.~Gromov, V.~Kazakov and P.~Vieira,
\textit{``{Exact Spectrum of Anomalous Dimensions of Planar $\mathcal{N}=4$
  Supersymmetric Yang-Mills Theory}''},
\textsf{\doiref{10.1103/PhysRevLett.103.131601}{Phys.~Rev.~Lett.~103,~131601~(%
2009)}},
\texttt{\arxivref{0901.3753}{arxiv:0901.3753}}.

\bibitem{Bombardelli:2009ns}
D.~Bombardelli, D.~Fioravanti and R.~Tateo,
\textit{``{Thermodynamic Bethe Ansatz for planar AdS/CFT: a proposal}''},
\textsf{\doiref{10.1088/1751-8113/42/37/375401}{J.~Phys.~A42,~375401~(2009)}},
\texttt{\arxivref{0902.3930}{arxiv:0902.3930}}.

\bibitem{Arutyunov:2007tc}
G.~Arutyunov and S.~Frolov,
\textit{``{On String S-matrix, Bound States and TBA}''},
\textsf{\doiref{10.1088/1126-6708/2007/12/024}{JHEP~0712,~024~(2007)}},
\texttt{\arxivref{0710.1568}{arxiv:0710.1568}}.

\bibitem{Arutyunov:2009zu}
G.~Arutyunov and S.~Frolov,
\textit{``{String hypothesis for the $AdS_5 \times S^5$ mirror}''},
\textsf{\doiref{10.1088/1126-6708/2009/03/152}{JHEP~0903,~152~(2009)}},
\texttt{\arxivref{0901.1417}{arxiv:0901.1417}}.

\bibitem{Bajnok:2008qj}
Z.~Bajnok, R.~A.~Janik and T.~Lukowski,
\textit{``{Four loop twist two, BFKL, wrapping and strings}''},
\textsf{\doiref{10.1016/j.nuclphysb.2009.02.005}{Nucl.~Phys.~B816,~376~(2009)}%
},
\texttt{\arxivref{0811.4448}{arxiv:0811.4448}}.

\bibitem{Bajnok:2009vm}
Z.~Bajnok, A.~Hegedus, R.~A.~Janik and T.~Lukowski,
\textit{``{Five loop Konishi from AdS/CFT}''},
\textsf{\doiref{10.1016/j.nuclphysb.2009.10.015}{Nucl.~Phys.~B827,~426~(2010)}%
},
\texttt{\arxivref{0906.4062}{arxiv:0906.4062}}.

\bibitem{Arutyunov:2010gb}
G.~Arutyunov, S.~Frolov and R.~Suzuki,
\textit{``{Five-loop Konishi from the Mirror TBA}''},
\textsf{\doiref{10.1007/JHEP04(2010)069}{JHEP~1004,~069~(2010)}},
\texttt{\arxivref{1002.1711}{arxiv:1002.1711}}.

\bibitem{Balog:2010xa}
J.~Balog and A.~Hegedus,
\textit{``{5-loop Konishi from linearized TBA and the XXX magnet}''},
\textsf{\doiref{10.1007/JHEP06(2010)080}{JHEP~1006,~080~(2010)}},
\texttt{\arxivref{1002.4142}{arxiv:1002.4142}}.

\bibitem{Lukowski:2009ce}
T.~Lukowski, A.~Rej and V.~N.~Velizhanin,
\textit{``{Five-Loop Anomalous Dimension of Twist-Two Operators}''},
\textsf{\doiref{10.1016/j.nuclphysb.2010.01.008}{Nucl.~Phys.~B831,~105~(2010)}%
},
\texttt{\arxivref{0912.1624}{arxiv:0912.1624}}.

\bibitem{Beccaria:2009eq}
M.~Beccaria, V.~Forini, T.~Lukowski and S.~Zieme,
\textit{``{Twist-three at five loops, Bethe Ansatz and wrapping}''},
\textsf{\doiref{10.1088/1126-6708/2009/03/129}{JHEP~0903,~129~(2009)}},
\texttt{\arxivref{0901.4864}{arxiv:0901.4864}}.

\bibitem{Velizhanin:2010cm}
V.~N.~Velizhanin,
\textit{``{Six-Loop Anomalous Dimension of Twist-Three Operators in
  $\mathcal{N}=4$ SYM}''},
\texttt{\arxivref{1003.4717}{arxiv:1003.4717}}.

\bibitem{Velizhanin:2008pc}
V.~N.~Velizhanin,
\textit{``{Leading transcedentality contributions to the four-loop universal
  anomalous dimension in $\mathcal{N}=4$ SYM}''},
\texttt{\arxivref{0811.0607}{arxiv:0811.0607}}.

\bibitem{Fiamberti:2008sm}
F.~Fiamberti, A.~Santambrogio, C.~Sieg and D.~Zanon,
\textit{``{Finite-size effects in the superconformal beta-deformed
  $\mathcal{N}=4$ SYM}''},
\textsf{\doiref{10.1088/1126-6708/2008/08/057}{JHEP~0808,~057~(2008)}},
\texttt{\arxivref{0806.2103}{arxiv:0806.2103}}.

\bibitem{Fiamberti:2008sn}
F.~Fiamberti, A.~Santambrogio, C.~Sieg and D.~Zanon,
\textit{``{Single impurity operators at critical wrapping order in the
  beta-deformed $\mathcal{N}=4$ SYM}''},
\textsf{\doiref{10.1088/1126-6708/2009/08/034}{JHEP~0908,~034~(2009)}},
\texttt{\arxivref{0811.4594}{arxiv:0811.4594}}.

\bibitem{Beccaria:2009hg}
M.~Beccaria and G.~F.~De Angelis,
\textit{``{On the wrapping correction to single magnon energy in twisted N=4 SYM}''},
\textsf{ \doiref{10.1142/S0217751X09047375}}{Int.~J.~Mod.~Phys.~A24~5803~(2009)}
\texttt{\arxivref{0903.0778}{arxiv:0903.0778}}.

\bibitem{Gromov:2010dy}
N.~Gromov and F.~Levkovich-Maslyuk,
\textit{``{Y-system and beta-deformed $\mathcal{N}=4$ Super-Yang-Mills}''},
\texttt{\arxivref{1006.5438}{arxiv:1006.5438}}.

\bibitem{Arutyunov:2010gu}
G.~Arutyunov, M.~de Leeuw and S.~J.~van Tongeren,
\textit{``{Twisting the Mirror TBA}''},
\textsf{ \doiref{10.1007/JHEP02(2011)025}{JHEP~1102~025~(2011)}},
\texttt{\arxivref{1009.4118}{arxiv:1009.4118}}.

\bibitem{Chetyrkin:1980pr}
K.~G.~Chetyrkin, A.~L.~Kataev and F.~V.~Tkachov,
\textit{``{New Approach to Evaluation of Multiloop Feynman Integrals: The
  Gegenbauer Polynomial $x$ Space Technique}''},
\textsf{Nucl.~Phys.~B174,~345~(1980)}.

\bibitem{Kotikov:1995cw}
A.~V.~Kotikov,
\textit{``{The Gegenbauer Polynomial Technique: the evaluation of a class of
  Feynman diagrams}''},
\textsf{\doiref{10.1016/0370-2693(96)00226-2}{Phys.~Lett.~B375,~240~(1996)}},
\texttt{\arxivref{hep-ph/9512270}{hep-ph/9512270}}.

\bibitem{Broadhurst:1985vq}
D.~J.~Broadhurst,
\textit{``{Evaluation of a class of Feynman diagrams for all numbers of loops
  and dimensions}''},
\textsf{\doiref{10.1016/0370-2693(85)90340-5}{Phys.~Lett.~B164,~356~(1985)}}.

\end{thebibliography}

\end{fmffile}
\end{document}